\documentstyle[preprint,aps,psfig,epsfig]{revtex}
\tighten       

\begin{document} 
\def\eqn#1{Eq.$\,$#1}
\def\mb#1{\setbox0=\hbox{$#1$}\kern-.025em\copy0\kern-\wd0
\kern-0.05em\copy0\kern-\wd0\kern-.025em\raise.0233em\box0}
\draft
\preprint{}

\title{The spatial correlations in the velocities arising from a random
distribution of point vortices.  }
\author{Pierre-Henri Chavanis and Cl\'ement Sire}
\address{Laboratoire de Physique Quantique (UMR C5626 du CNRS), Universit\'e
Paul Sabatier\\ 
118 route de Narbonne, 31062 Toulouse Cedex 4, France\\
(chavanis{@}irsamc2.ups-tlse.fr \& clement{@}irsamc2.ups-tlse.fr)\\
Tel: +33-5-61558231 \qquad Fax:+33-5-61556065} 
\date{\today}
\maketitle

\begin{abstract}

This paper is devoted to a statistical analysis of the velocity fluctuations
arising from a random distribution of point vortices in two-dimensional
turbulence. Exact results are derived for the correlations in the velocities
occurring at two points separated by an arbitrary distance. We find that the
spatial correlation function decays extremely slowly  with the
distance. We discuss formal analogies with the statistics of the gravitational field
in stellar systems.

\end{abstract}

\eject

\section{Introduction}
\label{sec_introduction}

Recently, several papers have focused on the statistics of velocity and velocity gradients produced by a random distribution of point vortices in two dimensions \cite{min,jimenez,weiss,chukbar,kuvshinov,cs}. This problem was first considered by Min {\it et al.}  \cite{min} and independently by Jim\'enez \cite{jimenez}. They showed, using different methods, that the velocity p.d.f. are Gaussian (but with a slow convergence) while the distribution of velocity gradients follow a Cauchy law. Their theoretical results were confirmed by numerical simulations of point vortices \cite{min} and by Direct Navier Stokes simulations of 2D decaying turbulence when the flow becomes dominated by a large number of coherent vortices \cite{jimenez}. Additional simulations of point vortex systems were performed by Weiss {\it et al.} \cite{weiss} who emphasized the importance of vortex pairs in the dynamics and their role in the tail of the velocity p.d.f. They showed in particular that these pairs are responsible for anomalous diffusion and proposed to model the motion of the vortices by an Ornstein-Uhlenbeck stochastic process combined with L\'evy walks.

A formal analogy exists between the statistics of the velocity fluctuations due to a collection of point vortices and the statistics of the gravitational field produced by a random distribution of stars. The only difference, of great importance, is the space dimensionality $D=2$ instead of $D=3$. In a long series of papers, Chandrasekhar and von Neumann \cite{c0,cn1,cn2,c3,c4} analyzed in detail the  distribution of the gravitational field arising from a random distribution of stars. Their study was inspired by the work of  Holtsmark \cite{holtsmark} concerning the fluctuations of the electric field in a gas composed of simple ions and by the work of von Smoluchowski  \cite{smoluchowski} concerning the persistence of fluctuations in the  Brownian theory. Chandrasekhar and Von Neumann rederived the Holtsmark distribution for ${\bf F}$, the gravitational field, and determined many other statistical measures for the correlations of ${\bf F}$ or the joint distribution of ${\bf F}$ and $d{\bf F}/dt$. Their motivation was to derive the expression for the speed of fluctuations and the diffusion coefficient of stars in a purely stochastic framework. Likewise in the vortex problem, the formation of binary stars alters the results of the statistical analysis at large field strengths. The beautiful work of Chandrasekhar and von Neumann is an imposing ``tour de force'' and provides detailed mathematical methods for analyzing the statistics of fluctuations in physics and astronomy. 

The close connexion between stellar and vortex systems has been investigated by one of us in a series of papers \cite{chav96,csr96,chav98a,chav98b,quasi,chav00,chavkin} and it was natural to consider the extension of the Chandrasekhar-von Neumann analysis to the case of point vortices. Therefore,  in Ref.  \cite{cs} we analyzed in some details the statistical
features of the stochastic velocity field produced by a random distribution of
point vortices in 2D turbulence. We rederived the results of Min {\it et al.} \cite{min} and Jim\'enez \cite{jimenez} for the distribution of velocity ${\bf V}$ and acceleration ${\bf A}$ (closely related to the velocity gradients) using the methods of  Chandrasekhar and von Neumann. This formalism is very powerful and allowed us to obtain generalizations of previous works: we considered the distributions of the {vectors} ${\bf V}$ and ${\bf A}$ (not only the projection along an axis), we gave expressions for these distributions valid for an {arbitrary} number of vortices (for finite $N$ these distributions can be evaluated numerically) and we
extended the analysis to an arbitrary spectrum of circulations among the vortices and to the case of non singular vortex ``blobs''. We also considered for the first time the joint distribution of velocity and acceleration $W({\bf V},{\bf A})$, determined the typical duration of the velocity fluctuations $T({ V})$ and proposed an expression for the diffusion coefficient $D$ of point vortices thereby justifying the phenomenological result of Weiss {\it et al.} \cite{weiss}. These theoretical results were found to be in good agreement with  numerical simulations \cite{sire} and experiments
\cite{hansen} of 2D decaying turbulence. They are also of great importance to build up a rational kinetic theory of point vortices \cite{chav98a,chavkin}.

The motivation of the present paper is to characterize the spatial correlations in the velocities occurring at two points separated by an arbitrary distance. This problem is customary in
turbulence but, in general, exact results are difficult to obtain. It is an
interest of our  model to allow for nice analytical solutions. Like in Ref. \cite{cs}, we consider a collection of $N$ point vortices with
circulation $\gamma$ randomly distributed in a disk of radius $R$. We assume
that the vortices have a spatial Poisson distribution, i.e. their positions are
independent  and  uniformly distributed over the entire domain. We are
particularly interested in the ``thermodynamical limit'' in which the number of
vortices and the  size of the domain go to infinity ($N\rightarrow\infty$,
$R\rightarrow\infty$)  in such a way that the vortex density $n={N\over \pi
R^{2}}$ remains finite. In this limit, the Poisson distribution is shown to be stationary \cite{novikov}. Therefore, if the vortices are initially uniformly distributed they will remain uniformly distributed in average during all the subsequent evolution. This property has been checked numerically by Jim\'enez \cite{jimenez} using Direct Navier Stokes simulations of 2D decaying turbulence. In the statistical theory of point vortices initiated by Onsager \cite{onsager} and further developed by Joyce \& Montgomery \cite{joyce} and Lundgren \& Pointin  \cite{pointin}, the Poisson distribution corresponds to a structureless equilibrium state with inverse temperature $\beta=0$ (when the angular momentum 
is not conserved) or $\beta\rightarrow +\infty$ (when the angular momentum is conserved). Of course, more general initial conditions are possible and lead to equilibrium states with $\beta<0$ in which the vortices are clustered in ``macrovortices'' \cite{joyce,pointin}. The statistics of fluctuations remain the same in these more general situations \cite{min,chavkin} but the vortices are expected to experience a {\it systematic drift} \cite{chav98a} (Chavanis, 1998) in addition to their diffusive motion, due to the inhomogeneity of the vortex cloud. It is in this context that a kinetic theory of point vortices, consistent with Onsager approach, can be constructed \cite{chav98a,chavkin}. At equilibrium the drift balances the scattering and maintains nontrivial density distributions. We shall restrict ourselves, however, in the present article to the case of a uniform distribution of point vortices for which the drift cancels out. 

In section \ref{sec_vpdf}, we recall known results concerning the distribution of velocities $W({\bf V})$ occuring at a fixed point (equations (\ref{W2})(\ref{W3}) and (\ref{W4})). This is essentially to set the notations that will be used in the sequel. Since the velocity distribution created by point vortices  is intermediate between Gaussian and L\'evy laws, we shall call it the ``marginal Gaussian distribution''. We also consider the case of non singular ``vortex blobs'' with a core of size $a$. We write down the exact characteristic function of  $W({\bf V})$ valid for all velocities and all core sizes (equation (\ref{WC2})). For ``extended'' vortices, we prove that the velocity distribution is Gaussian (with no tail) in agreement with the numerical observations of Jim\'enez \cite{jimenez} and Bracco {\it et al.} \cite{bracco}. For $a=0$ we recover the point vortex limit for which the velocity distribution has a Gaussian core and an algebraic tail. For ``small'' non singular vortices ($a\rightarrow 0$ but $a\neq 0$), we argue that the form of our characteristic function can explain the occurence of almost {\it exponential tails} observed by Jim\'enez \cite{jimenez} and Bracco {\it et al.} \cite{bracco}.    

In section \ref{sec_deltaV} we analyze the joint distribution of  velocity and velocity gradient $W({\bf V},\delta {\bf V})$. In subsection \ref{sec_Cauchy}, we rederive the Cauchy distribution for $\delta {\bf V}$ \cite{min,jimenez} using the method of Chandrasekhar and von Neumann (equation (\ref{W8})). In subsection \ref{sec_blobs}, we generalize this method to the case of ``vortex blobs'' and find the explicit characteristic function of $W(\delta {\bf V})$ valid for all velocities and core sizes (equation (\ref{blobs2})). We find that the distribution is Cauchy for small fluctuations and Gaussian for large fluctuations. This is in agreement with the asymptotic behaviours found by Min {\it et al.}\cite{min}. It is likely that in between the distribution passes through an {\it exponential tail} as observed numerically in Ref. \cite{min}. In subsection \ref{sec_deltaVV}, we determine an exact expression for the moment $\langle \delta {\bf V}\rangle _{\bf V}$, the average value of $\delta {\bf V}$ for a given velocity ${\bf V}$ (equations (\ref{dVV16}) and (\ref{dVV17})). When $V\rightarrow +\infty$, our results have a clear physical meaning in the nearest neighbor approximation. In this approximation, we determine the $n$-th conditional moments $\langle (\delta {\bf V})^{n}\rangle _{\bf V}$ of the velocity gradients.     

In section \ref{sec_dV}, we turn to the more difficult problem concerning the velocity correlations between two points separated by a finite distance. Special attention is devoted to the bivariate distribution $W({\bf V}_{0},{\bf V}_{1})$ that a velocity ${\bf V}_{0}$ occurs at $O$ and that, simultaneously, a velocity ${\bf V}_{1}$ occurs at a point separated from the first by a distance ${\bf r}_{1}$.
In subsection \ref{sec_V1V0}, we determine an exact but complicated expression for the moment $\langle {\bf V}_{1}\rangle _{{\bf V}_{0}}$ (equations (\ref{M4})(\ref{Z3}) and (\ref{Z4})).

More explicit results are obtained in section \ref{sec_V1average} when this quantity is averaged over all mutual orientations of ${\bf r}_{1}$ and ${\bf V}_{0}$. An exact expression is found for $\langle {\bf V}_{1}\rangle _{{V}_{0}}$ in subsection \ref{sec_V1V0mod} (equations (\ref{M14})(\ref{M16})). In subsection \ref{sec_be}, we give the asymptotic behaviour of this quantity when  $r_{1}\rightarrow +\infty$ (equations (\ref{M21})(\ref{M22})). 

In section \ref{sec_V1paraverage}, we consider the projection of $\langle {\bf V}_{1}\rangle _{{\bf V}_{0}}$ in the direction of ${\bf V}_{0}$ and its average over all mutual orientations of ${\bf r}_{1}$ and ${\bf V}_{0}$. This defines a quantity $\langle V_{1||}\rangle_{V_{0}}$ (subsection \ref{sec_V1parmoy}) whose expression is given by equation (\ref{M27}). In subsection \ref{sec_asy}, we give the asymptotic behaviours of this quantity  in the limits $r_{1}\rightarrow 0$ (equation (\ref{M31})) and  $r_{1}\rightarrow +\infty$ (equations (\ref{M34})(\ref{M35})). When $\langle V_{1||}\rangle_{V_{0}}$ is further averaged over all
possible values of ${V}_{0}$, this determines a function of $r_{1}$ alone which
characterizes the correlations in the velocities occurring simultaneously 
at two
points separated by a distance ${\bf r}_{1}$ (subsection \ref{sec_V1perpr1}). The exact expression of $\langle V_{1||}\rangle$ is given by equation (\ref{M50}) and its asymptotic behaviours for $r_{1}\rightarrow 0$ and  $r_{1}\rightarrow +\infty$ by equations (\ref{M51})(\ref{M52}). In subsection \ref{sec_foncnorm} we calculate the correlation function $\langle {\bf V}_{0}{\bf V}_{1}/V_{0}^{2}\rangle$. Its general expression is given by equation (\ref{norm4}) and its  asymptotic behaviours by equations (\ref{norm10})(\ref{norm11}).    

Finally, in section \ref{sec_corrfonction}, we calculate the spatial velocity autocorrelation function and the energy spectrum of point vortices. We find in subsection \ref{sec_fonc} that the spatial velocity correlation function decays extremely slowly with the distance (equation (\ref{N6})). The simple result $\langle {\bf V}_{0}{\bf V}_{1}\rangle={n\gamma^{2}\over 2\pi}\ln (R/r_{1})$, also derived more directly in Appendix \ref{sec_alt}, does not seem to have been given previously. In subsection \ref{sec_spectrum} we use this correlation function to determine the energy spectrum $E(k)$ of point vortices (equation (\ref{S7})). This offers an alternative to the method considered by Novikov \cite{novikov}. When $k\rightarrow +\infty$, we recover the classical result $E(k)\sim k^{-1}$ of Novikov and for  $k\rightarrow 0$ we find $E(k)\sim k$.   

Note that the spatial velocity autocorrelation function $\langle {\bf V}_{0}{\bf V}_{1}\rangle$ diverges logarithmically at small separations since the variance of the velocity is not defined. This is the reason why we consider more regular quantities such as $\langle V_{1||}\rangle\equiv \langle {\bf V}_{0}{\bf V}_{1}/{V}_{0}\rangle$  and $\langle {\bf V}_{0}{\bf V}_{1}/{V}_{0}^{2}\rangle$ that are finite at small separations. The computation of these quantities is difficult and is a important part of our work.

\section{The statistics of velocity fluctuations occurring at a fixed point}
\label{sec_vpdf}

We recall in this section the results concerning the distribution of velocity
fluctuations occuring at a fixed point. This problem was first considered by Min {\it et al.} \cite{min} and Jim\'enez \cite{jimenez} and later on by Weiss {\it et al.} \cite{weiss}, Chukbar \cite{chukbar}, Kuvshinov \& Schep \cite{kuvshinov} and Chavanis \& Sire  \cite{cs}. Min {\it et al.}  and Weiss {\it et al.} focus on the mathematical limit $N\rightarrow +\infty$ and claim that the velocity distribution is Gaussian by invoking a generalization of the Central Limit Theorem due to Ibragimov \& Linnik  \cite{ibragimov} (the proof is not straightforward because the variance of the velocity created by a single vortex diverges logarithmically). However, they stress that  the convergence to the Gaussian is extremely slow and that this slow convergence is responsible for discrepencies at large velocities. This is confirmed by direct numerical simulations of point vortices \cite{min,weiss}. Jim\'enez \cite{jimenez}  and  Chavanis \& Sire \cite{cs} adopt another point of view  by treating $N$ as a large number but not $\ln N$. This seems to be more appropriate to physical situations where the typical number of point vortices does not exceed $10^{4}$. They proved in that limit that the velocity p.d.f. has a Gaussian core and an algebraic tail (see equations (\ref{W2})(\ref{W3}) and (\ref{W4})). When $\ln N\rightarrow +\infty$, the tail is rejected to infinity and the distribution is ``purely'' Gaussian  in agreement with the theorem of Ibragimov and Linnik. All authors understood that the algebraic tail is produced by the nearest neighbor and found the correct exponent by a phenomenological approach. A mathematical justification of this result was given by Jim\'enez \cite{jimenez}  and  Chavanis \& Sire \cite{cs}  who explicitly derived the characteristic function (\ref{C2}) associated with the velocity distribution. The $\rho^{2}$ factor in the characteristic function  gives rise to the Gaussian core while the term in $\ln\rho$ gives rise to the algebraic tail  \cite{jimenez,cs}. This last result is not straightforward and the rigorous proof requires to evaluate the integrals in the complex plane as done by Chavanis \& Sire \cite{cs}. The characteristic function (\ref{C2}) was also determined by  Chukbar  \cite{chukbar} but he arbitrarily replaced the logarithmic factor appearing in (\ref{C2}) by a constant which is not justified. Therefore, he could not derive the algebraic tail.  Kuvshinov \& Schep \cite{kuvshinov} (2000) wrote the right distribution but justified the algebraic tail by the phenomenological nearest neighbor argument not by a rigorous calculation.

Below, we recall the main results concerning the velocity p.d.f. following the presentation and notations of Ref. \cite{cs}. The velocity ${\bf V}$ occurring at the center $O$ of the domain  is the 
sum of the velocities ${\bf \Phi}_{i}$ $(i=1,...,N)$ produced by
the $N$ vortices:
\begin{equation}
{\bf V}=\sum_{i=1}^{N}{\bf \Phi}_{i}
\label{V1}
\end{equation}
\begin{equation}
{\bf \Phi}_{i}=-{\gamma\over 2\pi}{{\bf r}_{ \perp i} \over r_{i}^{2}}  
\label{Phi1}
\end{equation}
where ${\bf r}_{i}$ denotes the position of the $i^{th}$ vortex and, by
definition,  ${\bf r}_{\perp i}$ is the vector ${\bf r}_i$ rotated by
$+{\pi\over 2}$. Since the vortices are randomly distributed, the velocity ${\bf
V}$ fluctuates. In the limit
when the number of vortices and the size of the domain go to infinity
$N,R\rightarrow\infty$ in such a way that the density $n={N\over \pi R^{2}}$
remains finite, the distribution of the velocity ${\bf V}$ is
given by \cite{cs} 
\begin{equation}
W({\bf V})={1\over 4\pi^{2}}\int A({\mb \rho})e^{-i {\mb \rho}{\bf
V}}d^{2}{\mb\rho} 
\label{W1}
\end{equation}
with
\begin{equation}
A({\mb \rho})= e^{-n C({\mb \rho})}
\label{A1}
\end{equation} 
and
\begin{equation}
C({\mb \rho})={\gamma^{2}\over 4\pi^{2}}\int_{|{\mb\Phi}|={\gamma\over 2\pi
R}}^{+\infty} (1-e^{i{\mb\rho}{\mb\Phi}}){1\over\Phi^{4}}d^{2}{\mb\Phi}
\label{C1}
\end{equation} 
Introducing polar coordinates and performing the angular integration, equation (\ref{C1}) can be reduced to 
\begin{equation}
C({\mb \rho})={\gamma^{2}\rho^{2}\over 2\pi}\int_{\gamma\rho\over 2\pi
R}^{+\infty} (1-J_{0}(x)){dx\over x^{3}}
\label{WC1}
\end{equation} 
Because of logarithmic divergences when $R\rightarrow +\infty$, the ``thermodynamical limit'' does not
exist. Therefore, (\ref{W1}) must be considered as an equivalent of $W_{N}({\bf
V})$  for large $N$'s not a true limit. In Ref. \cite{jimenez,cs}, it was found that
\begin{equation}
C({\mb \rho})={\gamma^{2}\rho^{2}\over 16\pi}\ln\biggl ({4\pi N\over
n\gamma^{2}\rho^{2}}\biggr )
\label{C2}
\end{equation} 
For $\rho> 0$ and $N\rightarrow\infty$, we have
\begin{equation}
C({\mb \rho})={\gamma^{2}\rho^{2}\over 16\pi}\ln N
\label{C2bis}
\end{equation} 
and for $\rho\rightarrow 0$, we obtain
\begin{equation}
C({\mb \rho})=-{\gamma^{2}\rho^{2}\over 8\pi}\ln \rho
\label{C2tris}
\end{equation} 
The velocity distribution (\ref{W1}) is just the Fourier transform of 
(\ref{A1}) with (\ref{C2bis})-(\ref{C2tris}). It is found that (see Ref. \cite{cs} for detailed calculations): 
\begin{equation}
W({\bf  V})= {4\over  {n\gamma^{2}} \ln N} e^{-{4\pi\over n {\gamma^{2}}\ln N}
{V}^{2}}\quad (V\lesssim V_{crit}(N))
\label{W2}
\end{equation} 
\begin{equation}
W({\bf V})\sim {n\gamma^{2}\over 4 \pi^{2}V^{4}}\quad (V\gtrsim V_{crit}(N))
\label{W3}
\end{equation}
where 
\begin{equation}
V_{crit}(N)\sim \biggl ({n\gamma^{2}\over 4\pi}\ln N\biggr )^{1/2}\ln^{1/2}(\ln
N)
\label{W4}
\end{equation}

As discussed in Ref. \cite{cs}, the velocity p.d.f behaves in a manner which is
intermediate between Gaussian and L\'evy laws. This is because the variance
$\langle \Phi^{2}\rangle$ of the velocity created by a single vortex diverges
logarithmically. Therefore, the Central Limit Theorem is only marginally
applicable: the core of the velocity distribution is Gaussian while the tail
decays algebraically as for a L\'evy law. Therefore, the distribution function  (\ref{W2})(\ref{W3}) can be called a ``marginal Gaussian distribution''. The velocity $V_{crit}(N)$ marks the transition between the Gaussian core and the algebraic tail. This  algebraic tail can be given a simple interpretation in the nearest neighbor approximation (see, e.g., Ref. \cite{cs}). It is remarkable that the velocity created by the nearest neighbor 
\begin{equation}
V_{n.n.}^{2}\sim \biggl ({\gamma\over 2\pi d}\biggr )^{2}\sim {\gamma^{2}\over
4\pi^{2}}{N\over \pi R^{2}}
\label{Vnn}
\end{equation}
is precisely of the same order (up to some logarithmic corrections) as the
typical velocity created by the rest of the system
\begin{equation}
V^{2}\sim N \biggl \langle {\gamma^{2}\over 4\pi^{2} r^{2}}\biggr \rangle\sim
N\int_{|{\bf r}|=d}^{R} \tau({\bf r}){\gamma^{2}\over 4\pi^{2}r^{2}}d^{2}{\bf
r}\sim {\gamma^{2}\over 4\pi}{N\over \pi R^{2}}\ln N
\label{Vtyp}
\end{equation}  
where $d\sim n^{-1/2}$ denotes the interparticle distance and $\tau({\bf
r})=1/\pi R^{2}$ the probability of occurrence of a vortex in ${\bf r}$. In a
sense, we can consider that the velocity is dominated by the contribution of the
nearest neighbor and that collective effects are responsible for logarithmic
corrections.

The variance of the velocity can be written as
\begin{equation}
\langle V^{2}\rangle= {N\over \pi R^{2}}\int_{0}^{+\infty} {\gamma^{2}\over
4\pi^{2}r^{2}}2\pi r dr 
\label{VarDir}
\end{equation}  
The integral (\ref{VarDir}) diverges logarithmically at both small and large separations. This is consistent
with the distribution (\ref{W2})(\ref{W3}) and the formula
\begin{equation}
\langle V^{2}\rangle=\int_{0}^{+\infty} W({\bf V})V^{2}2\pi V dV
\label{varph}
\end{equation}  
Because of the algebraic tail, the integral (\ref{varph}) diverges logarithmically as $V\rightarrow +\infty$ (i.e
$r\rightarrow 0$). In addition, if we were to extend the Gaussian distribution
(\ref{W2}) to all velocities we would find that the ``variance''
\begin{equation}
\langle V^{2}\rangle = {n\gamma^{2}\over 4\pi}\ln N
\label{var}
\end{equation}  
diverges logarithmically when $N\rightarrow +\infty$ (i.e $R\rightarrow
+\infty$).

The first moment of the velocity defined by
\begin{equation}
\langle |{\bf V}|\rangle=\int_{0}^{+\infty} W({\bf V})V 2\pi V dV
\label{firstph}
\end{equation}  
diverges logarithmically as
$N\rightarrow \infty$ but, unlike $\langle V^{2}\rangle$, converges for
$V\rightarrow +\infty$. To leading order in $\ln N$, we have:
\begin{equation}
\langle |{\bf V}|\rangle = \biggl ({n\gamma^{2}\over 16}\ln N\biggr )^{1\over 2}
\label{Vmoy}
\end{equation}

If we account for a spectrum of circulations among the vortices, the previous
results are maintained with $\overline{\gamma^{2}}$ in place of $\gamma^{2}$
\cite{cs}. In particular, for a neutral system consisting in an equal number of
vortices with circulation $+\gamma$ and $-\gamma$, the results are unchanged.
This is to be expected as a vortex with circulation $-\gamma$ located in $-{\bf
r}$ produces the same velocity as a vortex with circulation $+\gamma$ located in
${\bf r}$. Since the vortices are randomly distributed with uniform probability,
the two populations are statistically equivalent. When the system is non 
neutral (but still homogeneous), the previous results derived at the center of the domain remain valid at any point provided that the velocity is replaced by the fluctuating velocity ${\mb {\cal V}}={\bf V}-\langle {\bf V}\rangle$ where $\langle {\bf V}\rangle={1\over 2}n\overline{\gamma}{\bf a}_{\perp}$ is the average velocity in ${\bf a}$ corresponding to a solid rotation \cite{cs}. The case of inhomogeneous distribution of point vortices is considered in \cite{chavkin}.

In reality, the vortices have a finite radius $a$ which is not necessarily small (vortex ``blobs''). If we consider that the core radius $a$ acts as a lower cut-off, equation (\ref{WC1}) must be replaced by 
\begin{equation}
C({\mb \rho})={\gamma^{2}\rho^{2}\over 2\pi}\int_{\gamma\rho\over 2\pi
R}^{\gamma\rho\over 2\pi a} (1-J_{0}(x)){dx\over x^{3}}
\label{WC2}
\end{equation} 
This expression can be further simplified by part integrations. When $a$ is sufficiently large, we can make the approximation  
\begin{equation}
C({\mb \rho})\simeq {\gamma^{2}\rho^{2}\over 8\pi}\ln\biggl ({R\over a}\biggr )
\label{WC2app}
\end{equation} 
and this proves that for ``extended vortices'' the velocity distribution is exactly  Gaussian (with no tail) \cite{cs}. This result is consistent with the numerical observations of Jim\'enez \cite{jimenez} and Bracco {\it et al.}\cite{bracco}. For $a=0$, we recover the marginal Gaussian distribution  (\ref{W2})(\ref{W3}) with an algebraic tail. For ``small'' non singular vortices ($a\rightarrow 0$ but $a\neq 0$), the characteristic function (\ref{WC2}) is quadratic for $\rho\rightarrow +\infty$ and $\rho\rightarrow 0$ corresponding to a Gaussian velocity distribution for $V\rightarrow 0$ and $V\rightarrow +\infty$ (this large velocity limit is purely formal since, physically, the distribution must be cut at the maximum allowable velocity $V_{max}\sim \gamma/4\pi a$ achieved when two vortices are at distance $\sim 2a$ from each other). However, for intermediate values of $\rho$, $C({\mb \rho})$ is {\it not} quadratic and depends on the value of $a$. This fact can probably explain the occurence of almost {\it exponential tails}  observed by Jim\'enez  and Bracco {\it et al.} in their simulations.

\section{The velocity correlations between two neighboring points}
\label{sec_deltaV}

In this section, we analyze the spatial correlations of the velocity fluctuations. We
first consider the correlations between two neighboring points separated by an
infinitesimal distance $\delta {\bf r}$.

\subsection{The formal solution of the problem}
\label{sec_formaldelta}

The difference between the velocities occurring at two points distant $\delta
{\bf r}$ from each other is given by:
\begin{equation}
\delta {\bf V}=\sum_{i=1}^{N}{\mb \psi}_{i} 
\label{deltaV1}
\end{equation} 
with
\begin{equation} 
{\mb \psi}_{i} =-{\gamma\over 2\pi}\biggl \lbrace {\delta {\bf r}_{\perp}\over
r_{i}^{2}}-{2 ({\bf r}_{i}\delta{\bf r}){\bf r}_{i\perp}\over r_{i}^{4}}\biggr
\rbrace
\label{Psi1}
\end{equation}    
where we have assumed that one of the points is at the center of the domain.
This is similar to the expression for the acceleration ${\bf A}$ with
$\delta{\bf r}$ replacing ${\bf v}$ (see Ref. \cite{cs}, equations (4)(5)). The
correlations in the velocities occurring between these points can be specified
by the function $W({\bf V},\delta {\bf V})$ which gives the simultaneous
probability of the velocity ${\bf V}$ and the velocity increment $\delta {\bf
V}$. A general expression for the bivariate probability
$W({\bf V},\delta {\bf V})$ can be readily written down following Markov's
method outlined in Ref. \cite{cs}, section II.A. We have: 
\begin{equation}
W_{N}({\bf V},\delta {\bf V})={1\over 16\pi^{4}}\int A_{N}({\mb \rho},{\mb
\sigma}) e^{-i({\mb \rho}{\bf V}+{\mb \sigma}\delta {\bf V})}d^{2}{\mb
\rho}d^{2}{\mb \sigma}
\label{W5}
\end{equation}   
with:
\begin{equation}
A_{N}({\mb \rho},{\mb \sigma})=\biggl (\int_{|{\bf r}|=0}^{R} \tau({\bf r})
e^{i({\mb \rho}{\mb \Phi}+{\mb \sigma}{\mb \psi})}d^{2}{\bf r}\biggr )^{N}
\label{A4}
\end{equation}   
and
\begin{equation}
{\mb \Phi}=-{\gamma\over 2\pi}{{\bf r}_{\perp}\over r^{2}}
\label{Phi2}
\end{equation}
\begin{equation}
{\mb \psi}=-{\gamma\over 2\pi}\biggl ( {\delta {\bf r}_{\perp}\over r^{2}}-{2
({\bf r}\delta{\bf r}){\bf r}_{\perp}\over r^{4}}\biggr ) 
\label{Psi2}
\end{equation}
If we now suppose that the vortices are uniformly distributed on average, then
\begin{equation}
\tau({\bf r})={1\over\pi R^{2}}
\label{tau}
\end{equation}
and equation (\ref{A4}) becomes
\begin{equation}
A_{N}({\mb \rho},{\mb \sigma})=\biggl ({1\over \pi R^{2}}\int_{|{\bf r}|=0}^{R}
e^{i({\mb \rho}{\mb \Phi}+{\mb \sigma}{\mb \psi})}d^{2}{\bf r}\biggr )^{N}
\label{A5}
\end{equation} 
Since
\begin{equation}
{1\over \pi R^{2}} \int_{|{\bf r}|=0}^{R} d^{2}{\bf r}=1
\label{norm1}
\end{equation}
we can rewrite our expression for  $A_{N}({\mb \rho},{\mb \sigma})$ in the form
\begin{equation}
A_{N}({\mb \rho},{\mb \sigma})=\biggl (1-{1\over \pi R^{2}}\int_{|{\bf
r}|=0}^{R}(1-e^{i({\mb \rho}{\mb \Phi}+{\mb \sigma}{\mb \psi})})d^{2}{\bf
r}\biggr )^{N}
\label{A6}
\end{equation}   
We now consider the limit when the number of vortices and the size of the domain
go to infinity in such a way that the density remains finite
$$N\rightarrow\infty,\qquad R\rightarrow \infty,\qquad n={N\over \pi R^{2}}\quad
{\rm finite}$$ If the integral occurring in equation (\ref{A6}) increases less
rapidly than $N$, then
\begin{equation}
A({\mb \rho},{\mb \sigma})=e^{-n C({\mb \rho},{\mb \sigma})}
\label{A7}
\end{equation}   
with 
\begin{equation}
C({\mb \rho},{\mb \sigma})=\int_{|{\bf r}|=0}^{R}(1-e^{i({\mb \rho}{\mb
\Phi}+{\mb \sigma}{\mb \psi})})d^{2}{\bf r}
\label{C3}
\end{equation}
To make progress in the evaluation of (\ref{C3}), we find it more convenient to
take ${\mb \Phi}$ as a variable of integration instead of ${\bf r}$. The
Jacobian of the transformation $\lbrace{\bf r}\rbrace\rightarrow \lbrace{\bf
\Phi}\rbrace$ is
\begin{equation}
\biggl |\biggl | {\partial ({\bf r})\over\partial ({\bf \Phi})}\biggr |\biggr
|={\gamma^{2}\over 4\pi^{2}\Phi^{4}}
\label{Jac1}
\end{equation} 
so that
\begin{equation}
C({\mb \rho},{\mb \sigma})={\gamma^{2}\over 4\pi^{2}}\int_{|{\bf
\Phi}|={\gamma\over 2\pi R}}^{+\infty}(1-e^{i({\mb \rho}{\mb \Phi}+{\mb
\sigma}{\mb \psi})}){1\over \Phi^{4}}d^{2}{\bf \Phi}
\label{C4}
\end{equation}
In equation (\ref{C4}), ${\mb \psi}$  must be expressed in terms of ${\mb
\Phi}$. Combining (\ref{Phi2}) and (\ref{Psi2}) we find that
\begin{equation}
{\mb \psi}=-{2\pi\over \gamma}\biggl \lbrace \Phi^{2}\delta {\bf
r}_{\perp}+2({\mb \Phi}_{\perp}\delta {\bf r}){\mb \Phi}\biggr \rbrace
\label{Psi3}
\end{equation}

\subsection{The Cauchy distribution for $\delta{\bf V}$}
\label{sec_Cauchy}

In this subsection, we show that the unconditional distribution of the velocity fluctuations $\delta {\bf V}$ is governed by a 2D Cauchy law. This result was previously given by Jim\'enez \cite{jimenez} and Min {\it et al.}  \cite{min} using the methods of Ibragimov \& Linnik \cite{ibragimov} and Feller\cite{feller}. On a technical point of view, our calculations differ from the previous authors in the sense that we consider the distribution of the {\it vector}  $\delta {\bf V}$ instead of its projection along a particular direction (the same remark applies to the distribution of ${\bf V}$ in section \ref{sec_vpdf}). In addition, the method presented here can be extended to an arbitrary spectrum of circulations among the vortices and to the case of vortex ``blobs'' (subsection \ref{sec_blobs}).

According to equation (\ref{W5}), we clearly have
\begin{equation}
W(\delta {\bf V})={1\over 16\pi^{4}}\int A({\mb \rho},{\mb \sigma})
e^{-i({\mb \rho}{\bf V}+{\mb \sigma}\delta {\bf V})} d^{2}{\mb \rho}d^{2}{\mb
\sigma}d^{2}{\bf V}
\label{W6}
\end{equation} 
This gives the distribution of the velocity increment $\delta {\bf V}$ with no
condition on ${\bf V}$.  Using the identity
\begin{equation}
\delta({\bf x})={1\over (2\pi)^{2}}\int e^{-i{\mb\rho}{\bf x}}d^{2}{\mb\rho}
\label{delta1}
\end{equation}   
the foregoing expression for $W(\delta {\bf V})$ reduces to
\begin{equation}
W(\delta{\bf V})={1\over 4\pi^{2}}\int A({\mb\sigma}) e^{-i
{\mb\sigma}\delta{\bf V}} d^{2}{\mb\sigma}
\label{W7}
\end{equation}
where we have written $A({\mb\sigma})$ for $A({\bf 0},{\mb\sigma})$. Hence,
according to equations (\ref{A7}), (\ref{C4}) and (\ref{Psi3}) we obtain
\begin{equation}
A({\mb\sigma})=e^{-n C({\mb\sigma})}
\label{A8}
\end{equation}
with
\begin{equation}
C({\mb\sigma})={\gamma^{2}\over 4\pi^{2}}\int_{|{\mb\Phi}|=0}^{+\infty} (1-e^{i
{\mb\sigma}{\mb\psi}}){1\over\Phi^{4}}d^{2}{\mb\Phi}
\label{C5}
\end{equation}
and
\begin{equation}
{\mb\sigma}{\mb \psi}=-{2\pi\over \gamma}\biggl \lbrace
\Phi^{2}{\mb\sigma}\delta {\bf r}_{\perp}+2({\mb \Phi}_{\perp}\delta {\bf
r})({\mb \Phi}{\mb\sigma})\biggr \rbrace
\label{sPsi1}
\end{equation}
Following the usual prescription, we have let $R\rightarrow \infty$ in 
(\ref{C5}) since the
integral is convergent when $\Phi\rightarrow 0$. To evaluate this integral, we
shall first introduce a system of coordinates with the $x$-axis in the direction
of $\delta{\bf r}$. Let us denote by $\theta$ and $\beta$ the angles that
${\mb\Phi}$ and ${\mb\sigma}$ form with $\delta{\bf r}$. Equation (\ref{sPsi1})
now becomes
\begin{equation}
{\mb\sigma}{\mb \psi}={2\pi\over\gamma}\sigma \Phi^{2} |\delta {\bf r}|
\sin(2\theta-\beta)
\label{sPsi2}
\end{equation}  
and we obtain
\begin{equation}
C({\mb\sigma})={\gamma^{2}\over 4\pi^{2}}\int_{0}^{2\pi}\int_{0}^{+\infty}\biggl
(1-\cos\biggl \lbrack {2\pi\over \gamma}\sigma\Phi^{2} |\delta {\bf r}|
\sin(2\theta)\biggr\rbrack\biggr ){1\over\Phi^{3}}d\Phi d\theta
\label{C6}
\end{equation}
With the identity
\begin{equation}
\int_{0}^{2\pi}\cos(x\sin(2\theta))d\theta =2\pi J_{0}(x)
\label{id1}
\end{equation}
the expression (\ref{C6}) for $C({\mb\sigma})$ reduces to
\begin{equation}
C({\mb\sigma})={\gamma^{2}\over 2\pi}\int_{0}^{+\infty}\biggl (1-J_{0}\biggl (
{2\pi\over \gamma}\sigma\Phi^{2} |\delta {\bf r}|\biggr )\biggr
){1\over\Phi^{3}}d\Phi 
\label{C7}
\end{equation}
With the change of variables $x={2\pi\over \gamma}\sigma |\delta {\bf
r}|\Phi^{2}$, we obtain
\begin{equation}
C({\mb\sigma})={1\over 2}\gamma \sigma |\delta {\bf r}|\int_{0}^{+\infty}
(1-J_{0}(x)){dx\over x^{2}} 
\label{C8}
\end{equation}
Integrating by parts and using the identities
\begin{equation}
J_{0}'(x)=-J_{1}(x)
\label{id15}
\end{equation}
and
\begin{equation}
\int_{0}^{+\infty}{J_{1}(x)\over x}dx=1
\label{id2}
\end{equation} 
we find
\begin{equation}
C({\mb\sigma})={1\over 2}\gamma |\delta {\bf r}|\sigma
\label{C9}
\end{equation}
Hence
\begin{equation}
A({\mb\sigma})=e^{-{n\gamma\over 2}|\delta {\bf r}|\sigma}
\label{A9}
\end{equation}
The distribution $W(\delta{\bf V})$ is just the Fourier transform of (\ref{A9}).
This is the 2D-Cauchy distribution
\begin{equation}
W(\delta{\bf V})={2\over \pi n^{2}\gamma^{2} |\delta {\bf r}|^{2}}{1\over\biggl
(1+{4|\delta {\bf V}|^{2}\over n^{2}\gamma^{2} |\delta {\bf r}|^{2}}\biggr
)^{3/2}}
\label{W8}
\end{equation}
 The tail of the Cauchy distribution decreases algebraically like
\begin{equation}
W(\delta{\bf V})\sim  {n\gamma  |\delta {\bf r}|\over 4\pi |\delta {\bf V}|^{3}}
\label{W9}
\end{equation}

The distribution of velocity increments  $W(\delta {\bf V})$ is the same as the
distribution of accelerations  $W({\bf A})$. The only difference is the
occurrence of the displacement $|\delta {\bf r}|$ in place of the average
velocity $\overline{|{\bf v}|}$ (see Ref. \cite{cs}, equation (B30)). 

The Cauchy distribution is a particular L\'evy law. Accordingly, the typical
velocity increment produced by all the vortices
\begin{equation}
\delta V^{2}\sim N\int_{|{\bf r}|=d}^{R}\tau({\bf r}){\gamma^{2}\over
4\pi^{2}}|\delta{\bf r}|^{2}{1\over r^{4}}d^{2}{\bf r}\sim {\gamma^{2}\over
4\pi}|\delta{\bf r}|^{2}\biggl ({N\over \pi R^{2}}\biggr )^{2}
\label{deltaVtyp}
\end{equation}
 is dominated by the contribution of the nearest neighbor
\begin{equation}
\delta V_{n.n.}^{2}\sim {\gamma^{2}\over 
4\pi^{2}} {|\delta{\bf r}|^{2}\over d^{4}}\sim  {\gamma^{2}\over
4\pi^{2}}|\delta{\bf r}|^{2}\biggl ({N\over \pi R^{2}}\biggr )^{2}
\label{deltaVnn}
\end{equation}
In addition, the algebraic tail (\ref{W9}) has a clear physical interpretation
in the nearest neighbor approximation (see, e.g., Ref. \cite{cs}, section V).

Accounting for a spectrum of circulations among the vortices and following a
procedure similar to that adopted in the Appendix B of Ref. \cite{cs}, we find that
the previous results are maintained provided that $\gamma$ is replaced by
$\overline{|\gamma|}$. In particular, for a neutral system consisting in an
equal number of vortices with circulation $+\gamma$ and $-\gamma$, the results
are unchanged. This is to be expected since a vortex with circulation $-\gamma$
located in ${\bf r}_{\perp}$ produces the same velocity increment as a vortex
with circulation $+\gamma$ located in ${\bf r}$.

\subsection{The distribution $W(\delta {\bf V})$ for vortex ``blobs''}
\label{sec_blobs}

If the vortices are not point-like but rather ``blobs'', their size $a$ acts as
a lower cut-off and equation (\ref{C7}) must be replaced by:
\begin{equation}
C({\mb\sigma})={\gamma^{2}\over 2\pi}\int_{0}^{{\gamma\over 2\pi a}}\biggl
(1-J_{0}\biggl ( {2\pi\over \gamma}\sigma\Phi^{2} |\delta {\bf r}|\biggr )\biggr
){1\over\Phi^{3}}d\Phi 
\label{blobs1}
\end{equation}
After integrating by parts, we obtain
\begin{equation}
C({\mb\sigma})=\pi a^{2}\biggl\lbrack J_{0}\biggl ({\gamma\over 2\pi
a^{2}}|\delta{\bf r}|\sigma\biggr )-1\biggr\rbrack+{1\over 2}\gamma |\delta{\bf
r}|H\biggl ({\gamma\over 2\pi a^{2}}\sigma|\delta {\bf r}|\biggr )\sigma
\label{blobs2}
\end{equation} 
where $H(x)$ denotes the function
\begin{equation}
H(x)=\int_{0}^{x} {J_{1}(t)\over t}dt
\label{blobs3}
\end{equation} 

For $\sigma\rightarrow 0$, the function $C({\mb\sigma})$ is quadratic:
\begin{equation}
C({\mb\sigma})\sim{\gamma^{2}\over 16\pi}{|\delta{\bf r}|^{2}\over
a^{2}}\sigma^{2}
\label{blobs4}
\end{equation} 
implying that $W(\delta {\bf V})$ is Gaussian for large velocity increments 
\begin{equation}
W(\delta{\bf V})\sim{4\over n\gamma^{2}}{a^{2}\over |\delta {\bf
r}|^{2}}e^{-{4\pi\over n\gamma^{2}}{a^{2}\over |\delta {\bf r}|^{2}}|\delta {\bf
V}|^{2}}\qquad (|\delta {\bf V}|\rightarrow +\infty)
\label{blobs5}
\end{equation} 
The variance of the distribution (\ref{blobs5})  is
\begin{equation}
\langle (\delta {\bf V})^{2}\rangle ={n\gamma^{2}\over 4\pi}{|\delta {\bf
r}|^{2}\over a^{2}}
\label{blobs6}
\end{equation} 
as can be seen directly from equation (\ref{deltaVtyp}) introducing a cut-off at
$r\sim a$ instead of $d$.

For $\sigma\rightarrow +\infty$, the function $C({\mb\sigma})$ is linear
\begin{equation}
C({\mb\sigma})\sim {1\over 2}\gamma |\delta {\bf r}|\sigma 
\label{blobs7}
\end{equation} 
and we recover the Cauchy distribution (\ref{W8}) for small velocity increments.
Therefore the distribution $W(\delta {\bf V})$ makes a smooth transition from
Cauchy (concave in a semi-log plot) for small fluctuations to Gaussian (convex
in a semi-log plot) for large fluctuations. These asymptotic limits were found by Min {\it et al.} \cite{min}, but the general expression for the characteristic function (\ref{blobs2}) is new. It is likely that the two distributions are connected by an {\it exponential tail} as predicted and observed numerically by
 Min {\it et al.} Of course, when $a$ is reduced, the
transition happens at larger fluctuations (see equation (\ref{blobs6})) and we have typically a Cauchy distribution. Inversely, for ``extended'' vortices (large $a$), the distribution of velocity increments tends to a Gaussian. Similar results were obtained in Ref. \cite{cs} for the distribution of accelerations (note, however, that the functional form of the 
function $C({\mb\sigma})$, equation (\ref{blobs2}), is not exactly the same).

\subsection{The moment $\langle \delta {\bf V}\rangle _{\bf V}$}
\label{sec_deltaVV}

The average value of $\delta {\bf V}$ for a given velocity ${\bf V}$ is given
by
\begin{equation}
\langle \delta {\bf V}\rangle _{\bf V}={1\over W({\bf V})}\int W({\bf V},\delta
{\bf V})\delta {\bf V}d^{2}(\delta {\bf V})
\label{dVV1}
\end{equation}
According to equation (\ref{W5}) it can be rewritten
\begin{equation}
W({\bf V})\langle \delta {\bf V}\rangle _{\bf V}={1\over (2\pi)^{4}}\int
A({\mb\rho},{\mb\sigma}) e^{-i({\mb \rho}{\bf V}+{\mb\sigma}\delta {\bf
V})}\delta {\bf V} d^{2}{\mb \rho}d^{2}{\mb\sigma}d^{2}(\delta {\bf V})
\label{dVV2}
\end{equation}
or, equivalently,
\begin{equation}
W({\bf V})\langle \delta {\bf V}\rangle _{\bf V}=i{1\over (2\pi)^{4}}\int
A({\mb\rho},{\mb\sigma}) {\partial\over\partial {\mb\sigma}}\biggl\lbrace
e^{-i({\mb \rho}{\bf V}+{\mb\sigma}\delta {\bf V})}\biggr\rbrace d^{2}{\mb
\rho}d^{2}{\mb\sigma}d^{2}(\delta {\bf V})
\label{dVV3}
\end{equation}
Integrating by parts, we obtain
\begin{equation}
W({\bf V})\langle \delta {\bf V}\rangle _{\bf V}=-i {1\over (2\pi)^{4}}\int
{\partial A\over \partial {\mb \sigma}}({\mb \rho},{\mb \sigma}) e^{-i ({\mb
\rho}{\bf V}+{\mb \sigma}{\delta {\bf V})}}d^{2}{\mb \rho}d^{2}{\mb
\sigma}d^{2}(\delta {\bf V})
\label{dVV4}
\end{equation}
Using the identity (\ref{delta1}), we can readily carry out  the integration on
$\delta {\bf V}$ and ${\mb\sigma}$ to finally get:
\begin{equation}
W({\bf V})\langle \delta {\bf V}\rangle _{\bf V}=-{i\over (2\pi)^{2}}\int
{\partial A\over \partial {\mb \sigma}}({\mb \rho},{\bf 0}) e^{-i {\mb \rho}{\bf
V}}d^{2}{\mb \rho}
\label{dVV5}
\end{equation} 
To go further in the evaluation of the integral, we need to determine the behaviour of $A({\mb \rho},{\mb \sigma})$ for $|{\mb
\sigma}| \rightarrow 0$. In Appendix \ref{sec_detailed1} it is found that
\begin{equation}
A({\mb\rho},{\mb\sigma})= e^{-n C({\mb\rho}) +i{{\gamma} n\over 2}|\delta{\bf
r}|\lbrace \sigma_{y}\cos(2\theta)-\sigma_{x}\sin(2\theta)\rbrace
+o(|{\mb\sigma}|^{2})} 
\label{A11}
\end{equation} 
where $(\sigma_{x},\sigma_{y})$ are the components of ${\mb\sigma}$ in a system of coordinates where the $x$-axis coincides with the direction of $\delta {\bf r}$ and $\theta$ denotes the angle that ${\mb\rho}$ forms with $\delta {\bf r}$. We have therefore 
\begin{equation}
{\partial A\over\partial\sigma_{x}}({\mb\rho},{\bf 0}) =- i {{\gamma} n\over
2}|\delta {\bf r}|\sin(2\theta)e^{-n C({\mb\rho})}
\label{A12}
\end{equation} 
\begin{equation}
{\partial A\over\partial\sigma_{y}}({\mb\rho},{\bf 0})=  i {{\gamma} n\over
2}|\delta {\bf r}|\cos(2\theta)e^{-n C({\mb\rho})}
\label{A13}
\end{equation} 
The $x$-component of equation (\ref{dVV5}) now becomes
\begin{equation}
W({\bf V})\langle \delta V_{x}\rangle_{{\bf V}}=-{{\gamma} n\over
8\pi^{2}}|\delta {\bf r}|\int_{0}^{2\pi}d\theta\int_{0}^{+\infty}\rho d\rho \cos
\lbrace \rho V\cos (\chi-\theta)\rbrace \sin(2\theta)e^{-n C(\rho)}
\label{dVV6}
\end{equation}
where $\chi$ denotes the angle that ${\bf V}$ forms with $\delta{\bf r}$. Using
the  expansion formula (\ref{id2bis}) and the identities (\ref{id5})(\ref{id6})
we can carry out the angular integration to finally obtain
\begin{equation}
W({\bf V})\langle \delta V_{x}\rangle_{{\bf V}}={{\gamma} n\over 4\pi}|\delta
{\bf r}|\sin(2\chi)\int_{0}^{+\infty} J_{2}(\rho V)e^{-n C(\rho)}\rho d\rho
\label{dVV7}
\end{equation}
For $V\lesssim V_{crit}(N)$, the contribution of small $\rho$'s in the integral
(\ref{dVV7}) is negligible and we can replace $C(\rho)$ by its approximate value
(\ref{C2bis}). Therefore 
\begin{equation}
W({\bf V})\langle \delta V_{x}\rangle_{{\bf V}}={{\gamma} n\over 4\pi}|\delta
{\bf r}|\sin(2\chi)\int_{0}^{+\infty} J_{2}(\rho V)e^{-{n\gamma^{2}\over
16\pi}\ln N\rho^{2}}\rho d\rho
\label{dVV8}
\end{equation}
Using (\ref{W2}) and the identity
\begin{equation}
\int_{0}^{+\infty} J_{2}(x)e^{-\alpha x^{2}} x dx=2-2\biggl ( 1+{1\over
4\alpha}\biggr ) e^{-{1\over 4\alpha}}
\label{id7}
\end{equation}
we obtain 
\begin{equation}
\langle \delta V_{x}\rangle_{{\bf V}}={n\gamma\over 2}B\biggl ({4\pi V^{2}\over
n\gamma^{2}\ln N}\biggr )|\delta{\bf r}|\sin(2\chi)\quad (V\lesssim V_{crit}(N))
\label{dVV9}
\end{equation}
where $B(x)$ denotes the function
\begin{equation}
B(x)={1\over x}(e^{x}-1-x)
\label{B1}
\end{equation}
Similarly, we find
\begin{equation}
\langle \delta V_{y}\rangle_{{\bf V}}=-{n\gamma\over 2}B\biggl ({4\pi V^{2}\over
n\gamma^{2}\ln N}\biggr )   |\delta{\bf r}|\cos(2\chi) \quad (V\lesssim
V_{crit}(N))
\label{dVV10}
\end{equation}

For $V\gtrsim V_{crit}(N)$, the integral (\ref{dVV7}) is dominated by small
values of $\rho$ and one must use the general expression (\ref{C2}) for
$C(\rho)$. With the change of variables $z=\rho V$, equation (\ref{dVV7})
becomes
\begin{equation}
W({\bf V})\langle \delta V_{x}\rangle_{{\bf V}}={{\gamma} n\over 4\pi
V^{2}}|\delta {\bf r}|\sin(2\chi)\int_{0}^{+\infty} J_{2}(z)e^{-n C({z\over V})}
z dz
\label{dVV11}
\end{equation} 
Using the recursion formula
\begin{equation}
J_{2}(z)={2\over z}J_{1}(z)-J_{0}(z)
\label{id8}
\end{equation} 
we obtain
\begin{equation}
W({\bf V})\langle \delta V_{x}\rangle_{{\bf V}}={{\gamma} n\over 4\pi
V^{2}}|\delta {\bf r}|\sin(2\chi) \biggl \lbrace  \int_{0}^{+\infty} 2
J_{1}(z)e^{-n C({z\over V})} dz - \int_{0}^{+\infty} J_{0}(z)e^{-n C({z\over
V})} z dz\biggr\rbrace
\label{dVV12}
\end{equation} 
When $V\rightarrow\infty$, the first integral is convergent while the second,
equal to $2\pi V^{2}W({\bf V})$, decreases like $V^{-2}$ [see equation
(\ref{W3})]. Therefore, to leading order in $1/V$:
\begin{equation}
W({\bf V})\langle \delta V_{x}\rangle_{{\bf V}}\sim{{\gamma} n\over 2\pi
V^{2}}|\delta {\bf r}|\sin(2\chi)  \int_{0}^{+\infty}  J_{1}(z) dz
\label{dVV13}
\end{equation}  
With (\ref{W3}) and the identity
\begin{equation}
\int_{0}^{+\infty}  J_{n}(x) dx =1
\label{id9}
\end{equation}  
we find
\begin{equation}
\langle \delta V_{x}\rangle_{{\bf V}}= {2\pi V^{2}\over \gamma}|\delta {\bf r}|
\sin(2\chi) \quad (V\gtrsim V_{crit}(N))
\label{dVV14}
\end{equation}  
Similarly
\begin{equation}
\langle \delta V_{y}\rangle_{{\bf V}}= -{2\pi V^{2}\over \gamma}|\delta {\bf r}|
\cos(2\chi) \quad (V\gtrsim V_{crit}(N))
\label{dVV15}
\end{equation}  
Equations (\ref{dVV9}) (\ref{dVV10}) and (\ref{dVV14}) (\ref{dVV15}) can be
written more compactly in the form
\begin{equation}
\langle \delta {\bf V} \rangle_{{\bf V}}=-{n\gamma\over 2}B\biggl ({4\pi
V^{2}\over n\gamma^{2}\ln N}\biggr )    \biggl\lbrace \delta {\bf
r}_{\perp}+2{({\bf V}_{\perp}\delta {\bf r})\over V^{2}}{\bf V} \biggr \rbrace
\quad (V\lesssim V_{crit}(N))
\label{dVV16}
\end{equation}
\begin{equation}
\langle \delta {\bf V} \rangle_{{\bf V}}=-{2\pi\over \gamma}V^{2}\biggl\lbrace
\delta {\bf r}_{\perp}+2{({\bf V}_{\perp}\delta {\bf r})\over V^{2}}{\bf V}
\biggr \rbrace \quad (V\gtrsim V_{crit}(N))
\label{dVV17}
\end{equation}  
where $B(x)$ is defined by equation (\ref{B1}). Equations (\ref{dVV16})
(\ref{dVV17}) can be compared with equation (172) of Chandrasekhar \&
von Neumann \cite{cn2} for the increment of the gravitational field between two
neighboring points.    

Equation (\ref{dVV17}) has a clear physical interpretation in the nearest
neighbor approximation. In Ref. \cite{cs} it was shown that the high velocity tail of
the distribution $W({\bf V})$ is produced solely by the nearest neighbor. Now,
the velocity and the velocity difference are related to the position ${\bf r}$
of the nearest neighbor by [see equations (\ref{Phi1}) and (\ref{Psi1})]:
\begin{equation}
{\bf V}=-{\gamma\over 2\pi}{{\bf r}_{\perp}\over r^{2}}
\label{Vsole}
\end{equation}
and 
\begin{equation}
\delta {\bf V}=-{\gamma\over 2\pi}\biggl \lbrace {\delta {\bf r}_{\perp}\over
r^{2}}-{2 ({\bf r}\delta{\bf r}){\bf r}_{\perp}\over r^{4}}\biggr \rbrace 
\label{dVsole}
\end{equation}   
Eliminating ${\bf r}$ between these two expressions, we obtain
\begin{equation}
( \delta {\bf V} )_{{\bf V}}=-{2\pi\over \gamma}\biggl\lbrace \delta {\bf
r}_{\perp}V^{2}+2({\bf V}_{\perp}\delta {\bf r}){\bf V} \biggr \rbrace
\label{dVVsole}
\end{equation}  
which coincides with the expression (\ref{dVV17}) of $\langle \delta {\bf V}
\rangle_{{\bf V}}$ for large $V$'s. Note that
\begin{equation}
( \delta {\bf V} )^{2}_{{\bf V}}={4\pi^{2}\over \gamma^{2}}|\delta {\bf
r}|^{2} V^{4}
\label{moh2}
\end{equation}  
From (\ref{dVVsole}) and (\ref{moh2}) we can determine all conditional moments of $\delta {\bf V}$ in the nearest neighbor approximation (i.e., valid for large $V$'s).

If we account for a spectrum of circulations among the vortices, we have in
place of equations (\ref{dVV16})(\ref{dVV17}): 
\begin{equation}
\langle \delta {\bf V} \rangle_{{\bf V}}=-{n\overline{\gamma}\over 2}B\biggl
({4\pi V^{2}\over n\overline{\gamma^{2}}\ln N}\biggr )    \biggl\lbrace \delta
{\bf r}_{\perp}+2{({\bf V}_{\perp}\delta {\bf r})\over V^{2}}{\bf V} \biggr
\rbrace   \quad (V\lesssim V_{crit}(N))
\label{dVV16bis}
\end{equation}
\begin{equation}
\langle \delta {\bf V} \rangle_{{\bf V}}=-{2\pi\overline{\gamma}\over
\overline{\gamma^{2}}}V^{2}\biggl\lbrace \delta {\bf r}_{\perp}+2{({\bf
V}_{\perp}\delta {\bf r})\over V^{2}}{\bf V} \biggr \rbrace \quad (V\gtrsim
V_{crit}(N))
\label{dVV17bis}
\end{equation}  
In particular, for a neutral system $\langle \delta {\bf V} \rangle_{{\bf
V}}={\bf 0}$. This is to be expected (at least for a symmetrical distribution of
circulations) since a vortex with circulation $-\gamma$ located in $-{\bf r}$
produces the same velocity but an opposite velocity increment as a vortex with
circulation $+\gamma$ located in ${\bf r}$. 

Physically, we must realize that the calculations developed in this section are not valid
for large values of $|{\bf V}|$ (for a fixed $|\delta{\bf r}|$). Indeed, large
velocities are produced by a vortex very close to the origin. In that case,
$|{\bf V}|={\gamma\over 2\pi r}\rightarrow\infty$ while the velocity produced in
$\delta {\bf r}$ is bound to the value $|{\bf V}_{1}|\sim {\gamma\over 2\pi
|\delta {\bf r}|}$. Therefore, the velocity difference $\delta {\bf V}={\bf
V}-{\bf V}_{1}$ should behave like $|{\bf V}|$. Now, according to formula
(\ref{dVV17}), we have $|\delta {\bf V}|\sim V^{2}$. This inconsistency is
related to the fact that, when $|{\bf V}|\rightarrow +\infty$, the velocity
difference $\delta {\bf V}$ cannot be represented to any degree of accuracy by
the Taylor expansion (\ref{Psi1}) which assumes $|\delta {\bf r}|\ll |{\bf r}|$.
We expect therefore that the results of this section will be valid only for
velocities $|{\bf V}|< {\gamma\over 2\pi |\delta {\bf r}|}$. In view of these
remarks, it is necessary to consider now the more general problem of the
velocity correlations between two points separated by a finite distance.

\section{The correlations in the velocities occurring at two points separated by
a finite distance }
\label{sec_dV}

\subsection{A general formula for $W({\bf V}_{0},{\bf V}_{1})$}
\label{sec_WVV}

The general expression for $W_{N}({\bf V}_{0},{\bf V}_{1})$, the bivariate
probability that  a velocity
\begin{equation}
{\bf V}_{0}=-{\gamma\over 2\pi}\sum_{i=1}^{N}{{\bf r}_{\perp i}\over r_{i}^{2}}
\label{V2}
\end{equation} 
occurs at the center of the domain and that, simultaneously, a velocity 
\begin{equation}
{\bf V}_{1}=-{\gamma\over 2\pi}\sum_{i=1}^{N}{({\bf r}_{i}-{\bf
r}_{1})_{\perp}\over |{\bf r}_{i}-{\bf r}_{1}|^{2}}
\label{V3}
\end{equation}
occurs at a point $M_{1}$ separated from the first by a distance  ${\bf r}_{1}$ can be readily
written down following Markov's method outlined in Ref. \cite{cs}, section II.A. We
have
\begin{equation}
W_{N}({\bf V}_{0},{\bf V}_{1})={1\over (2\pi)^{4}}\int A_{N}({\mb \rho},{\mb
\sigma}) e^{-i({\mb \rho}{\bf V}_{0}+{\mb \sigma}{\bf V}_{1})} d^{2}{\mb
\rho}d^{2}{\mb \sigma}
\label{W10}
\end{equation}   
with:
\begin{equation}
A_{N}({\mb \rho},{\mb \sigma})=\biggl (1- {1\over \pi R^{2}}\int_{|{\bf
r}|=0}^{R}\bigl(1- e^{-i{\gamma\over 2\pi}({{\mb \rho}{\bf r}_{\perp}\over
r^{2}}+{{\mb \sigma}({\bf r}-{\bf r}_{1})_{\perp}\over |{\bf r}-{\bf
r}_{1}|^{2}})}\bigr)d^{2}{\bf r}\biggr )^{N}
\label{A14}
\end{equation}   
We now consider the limit when the number of vortices and the size of the domain
go to infinity in such a way that the density remains finite. If the integral
occurring in equation (\ref{A14}) increases less rapidly than $N$, then
\begin{equation}
A({\mb \rho},{\mb \sigma}) = e^{-n C({\mb \rho},{\mb \sigma})}
\label{A15}
\end{equation} 
with
\begin{equation}
C({\mb \rho},{\mb \sigma})=\int_{|{\bf r}|=0}^{R} (1-e^{-i{\gamma\over
2\pi}({{\mb \rho}{\bf r}_{\perp}\over r^{2}}+{{\mb \sigma}({\bf r}-{\bf
r}_{1})_{\perp}\over |{\bf r}-{\bf r}_{1}|^{2}})})d^{2}{\bf r}
\label{C11}
\end{equation} 
Writing
\begin{equation}
C({\mb \rho},{\mb \sigma})=C({\mb \rho})+D({\mb \rho},{\mb \sigma})
\label{C12}
\end{equation}
where $C({\mb\rho})=C({\mb \rho},{\bf 0})$ is given by equation (\ref{C2}), we have
\begin{equation}
D({\mb \rho},{\mb \sigma})=\int_{|{\bf r}|=0}^{R} e^{-i{\gamma\over 2\pi}{{\mb
\rho}{\bf r}_{\perp}\over r^{2}}}\biggl (1-e^{-i{\gamma\over 2\pi}{{\mb
\sigma}({\bf r}-{\bf r}_{1})_{\perp}\over |{\bf r}-{\bf r}_{1}|^{2}})}\biggr )
d^{2}{\bf r}
\label{D8}
\end{equation} 
Equations (\ref{W10}) (\ref{A15}) (\ref{C12}) and (\ref{D8})  formally solve the
problem but it does not seem possible to obtain an explicit expression for
$D({\mb \rho},{\mb \sigma})$. However, like in section \ref{sec_deltaVV}, if we
are interested only in the moments of ${\bf V}_{1}$ for a given value of ${\bf
V}_{0}$, we  need only the behaviour of $D({\mb \rho},{\mb \sigma})$ for $|{\mb
\sigma}|\rightarrow 0$. Expanding the exponential term  which appears under the
integral sign  in a power series in $|{\mb\sigma}|$, we obtain to first order
\begin{equation}
D({\mb \rho},{\mb \sigma})=D^{(1)}({\mb \rho},{\mb \sigma})+o(|{\mb\sigma}|^{2})
\label{D9}
\end{equation}
with
\begin{equation}
D^{(1)}({\mb \rho},{\mb \sigma})={i\gamma\over 2\pi}\int_{|{\bf r}|=0}^{R}
e^{-i{\gamma\over 2\pi}{{\mb \rho}{\bf r}_{\perp}\over r^{2}}}{{\mb \sigma}({\bf
r}-{\bf r}_{1})_{\perp}\over |{\bf r}-{\bf r}_{1}|^{2}} d^{2}{\bf r}
\label{D10}
\end{equation}
According to equations (\ref{A15}) (\ref{C12}) and (\ref{D9}),
$A({\mb\rho},{\mb\sigma})$ can be expressed as:
\begin{equation}
A({\mb\rho},{\mb\sigma})=e^{-n C({\mb\rho})-n D^{(1)}({\mb \rho},{\mb \sigma})
+o(|{\mb\sigma}|^{2})}
\label{A16}
\end{equation}
In Appendix \ref{sec_detailed2}, it is found that
\begin{eqnarray}
D^{(1)}({\mb\rho},{\mb\sigma})={\gamma^{2}\over 2\pi}\rho
(\sigma_{x}\sin\theta_{1}+\sigma_{y}\cos\theta_{1})\int_{0}^{+\infty}{dr\over
r}J_{2}\biggl ( {\gamma\rho\over 2\pi r}\biggr )\ln r_{>}\nonumber\\
-(\sigma_{x}\sin\theta_{1}+\sigma_{y}\cos\theta_{1})
\sum_{n=1}^{+\infty}A_{n}(\rho,r_{1})\cos
(n\theta_{1}) \nonumber\\
-(\sigma_{x}\cos\theta_{1}-\sigma_{y}\sin\theta_{1})
\sum_{n=1}^{+\infty}B_{n}(\rho,r_{1})\sin
(n\theta_{1})\nonumber\\ 
-{\gamma^{2}\over 4\pi}\rho \ln R
(\sigma_{x}\sin\theta_{1}+\sigma_{y}\cos\theta_{1})-i{\gamma\over 2}
r_{1}\sigma_{y}
\label{D23}
\end{eqnarray}
where $r_{>}$ (resp. $r_{<}$) is the larger (resp. smaller) of $(r,r_{1})$,  $A_{n}(\rho,r_{1})$ and $B_{n}(\rho,r_{1})$ are defined by equations
(\ref{An1}) and (\ref{Bn1}), $(\sigma_{x},\sigma_{y})$ are the components of
${\mb\sigma}$ in a system of coordinates where the $x$-axis is in the direction of ${\bf r}_{1}$ and  $-\theta_{1}$ is the angle that ${\mb\rho}_{\perp}$ forms with ${\bf r}_{1}$.

\subsection{The first moment $\langle{\bf V}_{1}\rangle_{{\bf V}_{0}}$}
\label{sec_V1V0}

The average value of ${\bf V}_{1}$ for a given ${\bf V}_{0}$ is given by:
\begin{equation}
\langle {\bf V}_{1}\rangle_{{\bf V}_{0}}={1\over W({\bf V}_{0})}\int W({\bf
V}_{0},{\bf V}_{1}){\bf V}_{1}d^{2}{\bf V}_{1}
\label{M1}
\end{equation}
By a procedure similar to that adopted in section \ref{sec_deltaVV}, we find
that an equivalent expression for  $\langle {\bf V}_{1}\rangle_{{\bf V}_{0}}$ is
\begin{equation}
W({\bf V}_{0})\langle {\bf V}_{1}\rangle_{{\bf V}_{0}}=-{i\over (2\pi)^{2}}\int
{\partial  A\over\partial {\mb\sigma}}({\mb\rho},{\bf 0})  e^{-i{\mb\rho}{\bf
V}_{0}}  d^{2}{\mb\rho}
\label{M2}
\end{equation} 
or, using equation (\ref{A16})
\begin{equation}
W({\bf V}_{0})\langle {\bf V}_{1}\rangle_{{\bf V}_{0}}={in\over (2\pi)^{2}} \int
{\partial D^{(1)}\over\partial {\mb\sigma}} ({\mb\rho})e^{-n C({\mb\rho})}
e^{-i{\mb\rho}{\bf V}_{0}} d^{2}{\mb\rho} 
\label{M3}
\end{equation} 
Introducing  polar coordinates, it can be written:
\begin{equation}
W({\bf V}_{0})\langle {\bf V}_{1}\rangle_{{\bf V}_{0}}={in\over (2\pi)^{2}}
\int_{0}^{+\infty} {\bf Z}(\rho,{\bf V}_{0}) e^{-n C(\rho)}\rho d\rho  
\label{M4}
\end{equation} 
where
\begin{equation}
{\bf Z}(\rho,{\bf V}_{0})=\int_{0}^{2\pi}{\partial D^{(1)}\over\partial
{\mb\sigma}}({\mb\rho}) e^{-i\rho V_{0}\cos\phi} d\phi  
\label{Z1}
\end{equation}
In equation (\ref{Z1}), $\phi$ denotes the angle that ${\mb\rho}$ forms with
${\bf V}_{0}$. It is related to $\chi$ and $-\theta_{1}$, the angles that ${\bf
V}_{0}$ and ${\mb\rho}_{\perp}$  form with ${\bf r}_{1}$ by:
\begin{equation}
\phi=-\chi-\theta_{1}-{\pi\over 2}
\label{angles1}
\end{equation}
Therefore, an alternative expression for ${\bf Z}(\rho,{\bf V}_{0})$ is:
\begin{equation}
{\bf Z}(\rho,{\bf V}_{0})=\int_{0}^{2\pi}{\partial D^{(1)}\over\partial
{\mb\sigma}}({\mb\rho})  e^{i\rho V_{0}\sin(\chi +\theta_{1})} d\theta_{1}  
\label{Z2}
\end{equation} 
Using (\ref{D23}) and  the identity
\begin{equation}
e^{i x\sin\theta}=J_{0}(x)+2\sum_{n=1}^{+\infty}J_{2n}(x)\cos
(2n\theta)+2i\sum_{n=0}^{+\infty}J_{2n+1}(x)\sin ((2n+1)\theta) 
\label{id12}
\end{equation} 
we find after lengthy calculations that
\begin{eqnarray}
Z_{x}(\rho,{\bf V}_{0})= i\gamma^{2}\rho J_{1}(\rho
V_{0})\cos\chi\int_{0}^{+\infty}{dr\over r}J_{2}\biggl ( {\gamma\rho\over 2\pi
r}\biggr )\ln r_{>}-i{\gamma^{2}\over 2}\rho J_{1}(\rho V_{0})\cos\chi\ln R
\nonumber\\ 
+\pi\sum_{n=0}^{+\infty}\lbrack
C_{2n-1}(\rho,r_{1})-D_{2n+1}(\rho,r_{1})\rbrack J_{2n}(\rho V_{0})\sin
(2n\chi)\nonumber\\ 
-i\pi\sum_{n=0}^{+\infty}\lbrack
C_{2n}(\rho,r_{1})-D_{2n+2}(\rho,r_{1})\rbrack J_{2n+1}(\rho V_{0})\cos
((2n+1)\chi)
\label{Z3}
\end{eqnarray}
and 
\begin{eqnarray}
Z_{y}(\rho,{\bf V}_{0})= i\gamma^{2}\rho J_{1}(\rho
V_{0})\sin\chi\int_{0}^{+\infty}{dr\over r}J_{2}\biggl ( {\gamma\rho\over 2\pi
r}\biggr )\ln r_{>}\nonumber\\
 -i{\gamma^{2}\over 2}\rho J_{1}(\rho
V_{0})\sin\chi\ln R-i\pi\gamma r_{1}J_{0}(\rho V_{0})\nonumber\\
-\pi\sum_{n=0}^{+\infty}\lbrack C_{2n-1}(\rho,r_{1})+D_{2n+1}(\rho,r_{1})\rbrack
J_{2n}(\rho V_{0})\cos (2n\chi)\nonumber\\
 -i\pi\sum_{n=0}^{+\infty}\lbrack
C_{2n}(\rho,r_{1})+D_{2n+2}(\rho,r_{1})\rbrack J_{2n+1}(\rho
V_{0})\sin((2n+1)\chi)
\label{Z4}
\end{eqnarray}
where we have introduced the notations:
\begin{eqnarray}
C_{n}(\rho,r_{1})=A_{n}(\rho,r_{1})+B_{n}(\rho,r_{1})
\label{Cn1}
\end{eqnarray}
\begin{eqnarray}
D_{n}(\rho,r_{1})=A_{n}(\rho,r_{1})-B_{n}(\rho,r_{1})
\label{Dn1}
\end{eqnarray}
for $n>0$ and $C_{n},D_{n}=0$ otherwise. Evaluating $C_{n}$ and $D_{n}$ as
defined in equations (\ref{Cn1}) (\ref{Dn1}), we find explicitly that:
\begin{eqnarray}
C_{n}(\rho,r_{1})={\gamma^{2}\over 2\pi}\rho {i^{n}\over
n}\int_{0}^{+\infty}{dr\over r}\biggl ({r_{<}\over r_{>}}\biggr )^{n}
J_{n+2}\biggl( {\gamma\rho\over 2\pi r}\biggr )
\label{Cn2}
\end{eqnarray}
\begin{eqnarray}
D_{n}(\rho,r_{1})={\gamma^{2}\over 2\pi}\rho {i^{n}\over
n}\int_{0}^{+\infty}{dr\over r}\biggl ({r_{<}\over r_{>}}\biggr )^{n}
J_{n-2}\biggl ( {\gamma\rho\over 2\pi r}\biggr )
\label{Dn2}
\end{eqnarray}
and we recall that $J_{-n}(z)=(-1)^{n}J_{n}(z)$.

\section{The average value of ${\bf V}_{1}$ and the  function $\langle {\bf
V}_{1}\rangle_{V_{0}}$ }
\label{sec_V1average}

\subsection{The average velocity $\langle {\bf V}_{1}\rangle$ }
\label{sec_V1moy}

The moment $\langle{\bf V}_{1}\rangle$ is simply obtained by averaging $\langle
{\bf V}_{1}\rangle_{{\bf V}_{0}}$ over all possible values of ${\bf V}_{0}$:
\begin{equation}
\langle {\bf V}_{1}\rangle=\int W({\bf V}_{0})\langle {\bf V}_{1}\rangle_{{\bf
V}_{0}}d^{2}{\bf V}_{0}
\label{M5}
\end{equation}
Substituting equation  (\ref{M4}) in equation (\ref{M5}) and introducing polar
coordinates, this can be rewritten:
\begin{eqnarray}
\langle {\bf V}_{1}\rangle ={i n\over
(2\pi)^{2}}\int_{0}^{2\pi}d\chi\int_{0}^{+\infty}V_{0}dV_{0}
\int_{0}^{+\infty}\rho d\rho e^{-n C(\rho)}{\bf Z}(\rho,V_{0},\chi)
\label{M6}
\end{eqnarray}
where $\chi$ denotes the angle that ${\bf V}_{0}$ forms with ${\bf r}_{1}$. When
the integration is performed over $\chi$, using equations (\ref{Z3})(\ref{Z4}),
it is seen that the components of $\langle {\bf V}_{1}\rangle$ reduce to: 
\begin{equation}
\langle V_{1x}\rangle=0
\label{M7}
\end{equation} 
and:
\begin{equation}
\langle {V}_{1y}\rangle =-i {n\over
2}\int_{0}^{+\infty}V_{0}dV_{0}\int_{0}^{+\infty}\rho d\rho e^{-n C(\rho)} (
i\gamma r_{1}+D_{1}(\rho,r_{1}))  J_{0}(\rho V_{0})
\label{M8}
\end{equation} 
where, according to (\ref{Dn2}):
\begin{equation}
D_{1}(\rho,r_{1})=-i{\gamma^{2}\over 2\pi}\rho \int_{0}^{+\infty}J_{1}\biggl
({\gamma\rho\over 2\pi r}\biggr ){r_{<}\over r_{>}}{dr\over r}
\label{Dn3}
\end{equation}
Under this form,  it is not possible to interchange the  order of integration in
(\ref{M8}). However, an alternative expression for $\langle {V}_{1y}\rangle$ can
be obtained along the following lines. Writing 
\begin{equation}
\langle {V}_{1y}\rangle
=\int_{0}^{+\infty}dV_{0} \int_{0}^{+\infty}\rho V_{0} J_{0}(\rho
V_{0})\psi(\rho) d\rho
\label{M9}
\end{equation} 
where
\begin{equation}
\psi(\rho)=-i {n\over 2}e^{-n C(\rho)} ( i\gamma r_{1}+D_{1}(\rho,r_{1})) 
\label{psirho1}
\end{equation} 
and integrating by parts with the identity
\begin{equation}
x J_{0}(x)={d\over dx}(xJ_{1}(x))
\label{id13}
\end{equation} 
we obtain:
\begin{equation}
\langle {V}_{1y}\rangle
=-\int_{0}^{+\infty}dV_{0}\int_{0}^{+\infty}\rho J_{1}(\rho V_{0})\psi'(\rho) 
d\rho 
\label{M10}
\end{equation} 
It is now possible to interchange the order of integration. Using formula
(\ref{id9}), we find
\begin{equation}
\langle
{V}_{1y}\rangle=-\int_{0}^{+\infty}\psi'(\rho)d\rho=\psi(0)-
\psi(+\infty)={n\over 2}\gamma r_{1}
\label{M11}
\end{equation} 
Hence
\begin{equation}
\langle {\bf V}_{1}\rangle={1\over 2}n\gamma {\bf r}_{1\perp}
\label{M11bis}
\end{equation} 
This result is of course to be expected. It corresponds to the solid rotation
produced by a uniform distribution of point vortices with density $n\gamma$.

\subsection{The function $\langle {\bf V}_{1}\rangle_{V_{0}}$}
\label{sec_V1V0mod}

A simple result is obtained when 
$\langle {\bf V}_{1}\rangle_{{\bf V}_{0}}$ is averaged over all 
mutual orientations of ${\bf r}_{1}$ 
and ${\bf V}_{0}$. Define 
\begin{equation}
\langle {\bf V}_{1}\rangle_{V_{0}}={1\over 2\pi}\int_{0}^{2\pi} \langle {\bf
V}_{1}\rangle_{{\bf V}_{0}}d\chi
\label{M12}
\end{equation}
The integral of $\langle {\bf V}_{1}\rangle_{V_{0}}$ over $V_{0}$ is just the
average velocity in ${\bf r}_{1}$:
\begin{equation}
\langle {\bf V}_{1}\rangle=\int_{0}^{+\infty}W({\bf V}_{0}) \langle {\bf
V}_{1}\rangle_{{V}_{0}} 2\pi V_{0}dV_{0}
\label{M13}
\end{equation}
Therefore, comparing equation (\ref{M13}) with equation (\ref{M6}) we directly
obtain
\begin{equation}
\langle {V}_{1x}\rangle_{V_{0}}=0
\label{M14}
\end{equation} 
\begin{equation}
W({\bf V}_{0})\langle {V}_{1y}\rangle_{V_{0}}
=-i {n\over 4\pi}\int_{0}^{+\infty}\rho d\rho e^{-n C(\rho)}
( i\gamma r_{1}+D_{1}(\rho,r_{1}))  J_{0}(\rho V_{0})
\label{M15}
\end{equation} 
Using (\ref{Dn3}), we can rewrite our expression for $\langle {V}_{1y}\rangle_{V_{0}}$
in the form
\begin{equation}
W({\bf V}_{0})\langle {V}_{1y}\rangle_{V_{0}} =- {n\gamma^{2}\over
8\pi^2}\int_{0}^{+\infty}\rho^{2} d\rho e^{-n C(\rho)} J_{0}(\rho V_{0})
R(\rho,r_{1})
\label{M16}
\end{equation} 
where $R(\rho,r_{1})$ denotes the function
\begin{equation}
R(\rho,r_{1})=-{2\pi r_{1}\over \gamma\rho}+\int_{0}^{+\infty}J_{1}\biggl
({\gamma\rho\over 2\pi r}\biggr ){r_{<}\over r_{>}}{dr\over r}
\label{R1}
\end{equation}
Remembering that the range of integration in equation (\ref{R1}) has to be
broken at ${r}_{1}$ with the prescription given at the end of section
\ref{sec_WVV}, the equation defining $R(\rho,r_{1})$ has explicitly the form 
\begin{equation}
R(\rho,r_{1})=-{2\pi r_{1}\over \gamma\rho}+{1\over r_{1}}\int_{0}^{r_{1}}
J_{1}\biggl ({\gamma\rho\over 2\pi r}\biggr )dr+r_{1}\int_{r_{1}}^{+\infty}
J_{1}\biggl ({\gamma\rho\over 2\pi r}\biggr ){dr\over r^{2}}
\label{R2}
\end{equation}
With the change of variables $z={\gamma\rho\over 2\pi r}$, it becomes
\begin{equation}
R(\rho,r_{1})=-{2\pi r_{1}\over \gamma\rho}+{2\pi r_{1}\over \gamma\rho}
\int_{0}^{{\gamma\rho\over 2\pi r_{1}}} J_{1}(z) dz+{\gamma\rho\over 2\pi
r_{1}}\int_{{\gamma\rho\over 2\pi r_{1}}}^{+\infty} {J_{1}(z)\over z^{2}} dz
\label{R3}
\end{equation}
Using the identity
\begin{equation}
{d\over dz}\biggl ({J_{n}(z)\over z^{n}}\biggr )=-{J_{n+1}(z)\over z^{n}}
\label{id14}
\end{equation} 
for $n=0$ and integrating by parts, we obtain:
\begin{equation}
R(\rho,r_{1})=-{2\pi r_{1}\over \gamma\rho}J_{0}\biggl ({\gamma\rho\over 2\pi
r_{1}}\biggr )+{\gamma\rho\over 2\pi r_{1}}\int_{{\gamma\rho\over 2\pi
r_{1}}}^{+\infty} {J_{1}(z)\over z^{2}} dz
\label{R4}
\end{equation}
From equation (\ref{R4}) it is readily found that
\begin{equation}
R(\rho,r_{1})\rightarrow 0 \qquad (r_{1}\rightarrow 0)
\label{R5}
\end{equation}
\begin{equation}
R(\rho,r_{1})=-{2\pi r_{1}\over \gamma\rho}+{\gamma\rho\over 4\pi
r_{1}}\ln\biggl ({2\pi r_{1}\over \gamma\rho}\biggr ) \qquad (r_{1}\rightarrow
+\infty)
\label{R6}
\end{equation}
where we recall that $J_{1}(z)\sim {z\over 2}$ for $z\rightarrow 0$.

\subsection{The asymptotic behaviour of $\langle {\bf V}_{1}\rangle_{V_{0}}$ for
$r_{1}\rightarrow +\infty$}
\label{sec_be}

Substituting for the asymptotic expansion of $R(\rho,r_{1})$ from equation
(\ref{R6}) in equation (\ref{M16}), we obtain for $r_{1}\rightarrow +\infty$:
\begin{eqnarray}
W({\bf V}_{0})\langle {V}_{1y}\rangle_{V_{0}} ={n\gamma r_{1}\over
4\pi}\int_{0}^{+\infty} e^{-n C(\rho)}  J_{0}(\rho V_{0}) \rho   d\rho
\nonumber\\ +{n\gamma^{3}\over 32 \pi^{3} r_{1}}\int_{0}^{+\infty}e^{-n C(\rho)}
J_{0}(\rho V_{0}) \ln\biggl ({\gamma\rho\over 2\pi r_{1}}\biggr ) \rho^3 d\rho
\label{M17}
\end{eqnarray} 
The first integral is equal to $2\pi W({\bf V}_{0})$ so the first term is just
the average velocity $\langle {V}_{1y}\rangle$ [see section \ref{sec_V1moy}].
Defining
\begin{equation}
\Delta\langle {\bf V}_{1}\rangle_{V_{0}}
=\langle {\bf V}_{1}\rangle_{V_{0}}-\langle {\bf V}_{1}\rangle
\label{M18}
\end{equation} 
we have
\begin{equation}
W({\bf V}_{0})\Delta \langle  {\bf V}_{1}\rangle_{V_{0}}={n\gamma^{3}\over 32
\pi^{3}} {{\bf r}_{1\perp}\over r_{1}^{2}}\int_{0}^{+\infty}e^{-n C(\rho)}
J_{0}(\rho V_{0})  \ln\biggl ({\gamma\rho\over 2\pi r_{1}}\biggr )\rho^3 d\rho
\label{M19}
\end{equation} 

For $V_{0}\lesssim V_{crit}(N)$, the integral is dominated by large values of
$\rho$ and  we can make the approximation
\begin{equation}
W({\bf V}_{0})\Delta \langle  {\bf V}_{1}\rangle_{V_{0}}\simeq -{n\gamma^{3}\over
64 \pi^{3}} {{\bf r}_{1\perp}\over r_{1}^{2}}\ln N
\int_{0}^{+\infty}e^{-{n\gamma^{2}\over 16\pi}\ln N\rho^{2}} J_{0}(\rho V_{0})
\rho^3 d\rho
\label{M20}
\end{equation} 
Using (\ref{W2}) and the identity
\begin{equation}
\int_{0}^{+\infty}e^{-\alpha x^{2}}J_{0}(\beta x)x^{3}dx={1\over 2
\alpha^{2}}\biggl (1-{\beta^{2}\over 4 \alpha}\biggr ) e^{-{\beta^{2}\over 4
\alpha}}
\label{id15bis}
\end{equation} 
we obtain
\begin{equation}
\Delta \langle  {\bf V}_{1}\rangle_{V_{0}} =-{\gamma\over 2\pi} {{\bf
r}_{1\perp}\over r_{1}^{2}}\biggl (1-{4\pi V_{0}^{2}\over n\gamma^{2}\ln
N}\biggr ) \qquad (V_{0}\lesssim V_{crit}(N))
\label{M21}
\end{equation} 

For $V_{0}\gtrsim V_{crit}(N)$, we can follow the procedure outlined in
Ref. \cite{cs}, section II.C.,  and we find after some calculations
\begin{equation}
\Delta \langle {\bf V}_{1}\rangle_{V_{0}}
={\gamma\over 2\pi} {{\bf r}_{1\perp}\over r_{1}^{2}} \qquad (V_{0}\gtrsim
V_{crit}(N))
\label{M22}
\end{equation}   

According to these equations 
\begin{equation}
\Delta \langle  {\bf V}_{1}\rangle_{V_{0}}
=-{\gamma\over 2\pi} {{\bf r}_{1\perp}\over r_{1}^{2}} \qquad (V_{0}\rightarrow 0)
\label{M23}
\end{equation} 
\begin{equation}
\Delta \langle  {\bf V}_{1}\rangle_{V_{0}}
={\gamma\over 2\pi} {{\bf r}_{1\perp}\over r_{1}^{2}} \qquad (V_{0}\rightarrow
+\infty)
\label{M24}
\end{equation} 
This is similar to the velocity produced by a fictitious point vortex located in
$O$ with circulation $+\gamma$ (when $V_{0}\rightarrow 0$)  and $-\gamma$ (when
$V_{0}\rightarrow +\infty$). It is not clear whether these results have a deeper
significance than is apparent at first sight. We observe that $\Delta \langle  {\bf
V}_{1}\rangle_{V_{0}}$ changes sign at
\begin{equation} 
V_{c}=\biggl ({n\gamma^{2}\over 4\pi}\ln N\biggr )^{1/2}
\label{Vc}
\end{equation}  
We can also check that
\begin{equation}
\int_{0}^{+\infty} W({\bf V}_{0})\Delta \langle  {\bf V}_{1}\rangle_{V_{0}} 2\pi
V_{0}dV_{0} =0 
\label{zero1}
\end{equation}
in agreement with (\ref{M11bis}).

\section{The first moment of ${\bf V}_{1}$ in the direction of ${\bf V}_{0}$ and
its average}
\label{sec_V1paraverage}

\subsection{The  moment $\langle V_{1 ||}\rangle_{{V}_{0}}$}
\label{sec_V1parmoy}

A quantity of physical interest is the average value of ${\bf V}_{1}$ in the
direction of ${\bf V}_{0}$. Let us define
\begin{equation}
V_{1 ||}={\bf V}_{1}{\mb 1}_{0}=V_{1x}\cos\chi+V_{1y}\sin\chi
\label{Vpar1}
\end{equation} 
where ${\mb 1}_{0}$ is the unit vector in the direction of ${\bf V}_{0}$. Then
\begin{equation}
\langle V_{1 ||}\rangle_{{\bf V}_{0}}=\langle V_{1x}\rangle_{{\bf
V}_{0}}\cos\chi+\langle V_{1y}\rangle_{{\bf V}_{0}}\sin\chi
\label{M25}
\end{equation} 
A relatively simple formula can be obtained when the foregoing expression for
$\langle V_{1 ||}\rangle_{{\bf V}_{0}}$ is averaged over all possible
orientations of ${\bf V}_{0}$. Define
\begin{equation}
\langle V_{1 ||}\rangle_{{V}_{0}}={1\over 2\pi}\int_{0}^{2\pi}\langle V_{1
||}\rangle_{{\bf V}_{0}}d\chi
\label{M26}
\end{equation} 
When equation (\ref{M4}) is substituted in equation (\ref{M26}) and the
integration is performed over $\chi$, only the logarithmic terms survive in the
series (\ref{Z3}) (\ref{Z4}). We are thus left with
\begin{equation}
W({\bf V}_{0})\langle V_{1 ||}\rangle_{V_{0}}=-{ n\gamma^{2}\over
4\pi^{2}}\int_{0}^{+\infty}\rho^{2}d\rho e^{-n C(\rho)}J_{1}(\rho
V_{0})Q(\rho,r_{1})
\label{M27}
\end{equation} 
where we have written
\begin{equation}
Q(\rho,r_{1})=\int_{0}^{+\infty}J_{2}\biggl ( {\gamma\rho\over 2\pi r}\biggr
)\ln r_{>}{dr\over r}-{1\over 2}\ln R
\label{Q1}
\end{equation}

With the usual interpretation of the notation $r_{>}$, see section \ref{sec_WVV}, the equation defining $Q(\rho,r_{1})$ has explicitly the form 
\begin{equation}
Q(\rho,r_{1})=\ln r_{1}\int_{0}^{r_{1}}J_{2}\biggl ( {\gamma\rho\over 2\pi
r}\biggr ){dr\over r}+\int_{r_{1}}^{+\infty}J_{2}\biggl ( {\gamma\rho\over 2\pi
r}\biggr )\ln r {dr\over r}   -{1\over 2}\ln R
\label{Q2}
\end{equation} 
With the change of variables $z={\gamma\rho\over 2\pi r}$, it becomes:
\begin{equation}
Q(\rho,r_{1})=\ln r_{1}\int_{{\gamma\rho\over 2\pi
r_1}}^{+\infty}J_{2}(z){dz\over z}+\int_{0}^{{\gamma\rho\over 2\pi
r_1}}J_{2}(z)\ln \biggl ({\gamma\rho\over 2\pi z}\biggr ){dz\over z}   -{1\over
2}\ln R
\label{Q3}
\end{equation}
Using the identity (\ref{id14}) with $n=1$ and integrating by parts, we obtain:
\begin{equation}
Q(\rho,r_{1})={1\over 2}\ln\biggl ({\gamma\rho\over 2\pi R}\biggr )-{1\over
2}\ln\epsilon -\int_{\epsilon}^{{\gamma\rho\over 2\pi r_1}}{J_{1}(z)\over
z^{2}}dz
\label{Q4}
\end{equation}
where it is understood that $\epsilon\rightarrow 0$.

From equation (\ref{Q4}), it is readily found that 
\begin{equation}
Q(\rho,r_{1})={1\over 2}\ln\biggl ({\gamma\rho\over 2\pi R}\biggr )\qquad
(r_{1}\rightarrow 0)
\label{Q5}
\end{equation}
\begin{equation}
Q(\rho,r_{1})=-{1\over 2}\ln\biggl ({R\over r_{1}}\biggr )\qquad
(r_{1}\rightarrow +\infty)
\label{Q6}
\end{equation}

\subsection{The asymptotic behaviour of $\langle V_{1 ||}\rangle_{V_{0}}$ for
$r_{1}\rightarrow 0$ and $r_{1}\rightarrow +\infty$ }
\label{sec_asy}

First considering the limit of $\langle V_{1 ||}\rangle_{V_{0}}$ for
$r_{1}\rightarrow 0$, we have according to (\ref{M27}) and (\ref{Q5})
\begin{equation}
W({\bf V}_{0})\langle V_{1 ||}\rangle_{V_{0}}=-{n\gamma^{2}\over
8\pi^{2}}\int_{0}^{+\infty}  e^{-nC(\rho)} J_{1}(\rho V_{0})\ln\biggl
({\gamma\rho\over 2\pi R}\biggr ) \rho^{2}    d\rho
\label{M28}
\end{equation}  
With the expression (\ref{C2}) for $C(\rho)$, the foregoing expression can be 
rewritten:
\begin{equation}
W({\bf V}_{0})\langle V_{1 ||}\rangle_{V_{0}}={n\over 2\pi}\int_{0}^{+\infty}
C'(\rho) e^{-nC(\rho)} J_{1}(\rho V_{0})\rho   d\rho
\label{M29}
\end{equation} 
Using (\ref{id13}) and integrating by parts, we find
\begin{equation}
W({\bf V}_{0})\langle V_{1 ||}\rangle_{V_{0}}={V_{0}\over
2\pi}\int_{0}^{+\infty} e^{-n C(\rho)} J_{0}(\rho V_{0})\rho d\rho
\label{M30}
\end{equation}  
The integral is just $2\pi W({\bf V}_{0})$. Hence
\begin{equation}
\langle V_{1 ||}\rangle_{V_{0}}=V_{0} \qquad (r_{1}\rightarrow 0)
\label{M31}
\end{equation}  
a result of course to be expected.

Considering now the behaviour of $\langle V_{1 ||}\rangle_{V_{0}}$ for
$r_{1}\rightarrow +\infty$, we have according to (\ref{M27}) and (\ref{Q6}):
\begin{equation}
W({\bf V}_{0})\langle V_{1 ||}\rangle_{V_{0}}={n\gamma^{2}\over
8\pi^{2}}\ln\biggl ({R\over r_{1}}\biggr ) \int_{0}^{+\infty}  e^{-n C(\rho)}
J_{1}(\rho V_{0})\rho^{2}d\rho
\label{M32}
\end{equation}  
For $V_{0}\lesssim V_{crit}(N)$, we can make the approximation
\begin{equation}
W({\bf V}_{0})\langle V_{1 ||}\rangle_{V_{0}}\simeq {n\gamma^{2}\over
8\pi^{2}}\ln\biggl ({R\over r_{1}}\biggr ) \int_{0}^{+\infty}
e^{-{n\gamma^{2}\over 16\pi}\ln N\rho^{2}} J_{1}(\rho V_{0})\rho^{2}d\rho
\label{M33}
\end{equation}
Integrating by parts, we obtain
\begin{equation}
\langle V_{1 ||}\rangle_{V_{0}}={2\over \ln N}V_{0}\ln\biggl ({R\over
r_{1}}\biggr ) \qquad (V_{0}\lesssim V_{crit}(N))
\label{M34}
\end{equation} 
For $V_{0}\gtrsim V_{crit}(N)$, we can adapt the procedure outlined in Ref. \cite{cs}
section II.C and we find after some calculations
\begin{equation}
\langle V_{1 ||}\rangle_{V_{0}}={n\gamma^{2}\over \pi V_{0}}\ln\biggl ({R\over
r_{1}}\biggr ) \qquad (V_{0}\gtrsim  V_{crit}(N))
\label{M35}
\end{equation}

\subsection{The moment $\langle V_{1 ||}\rangle$}  
\label{sec_V1perpr1}

If we now average $\langle V_{1 ||}\rangle_{V_{0}}$ over ${V}_{0}$, we shall
obtain a function of $|{\bf r}_{1}|$ only:
\begin{equation}
\langle V_{1 ||}\rangle=\int_{0}^{+\infty} W({\bf V}_{0})\langle V_{1
||}\rangle_{{ V}_{0}}2\pi V_{0} dV_{0}
\label{M36}
\end{equation} 
which will describe the correlations in the velocities occurring simultaneously
at two points separated by a distance  $|{\bf r}_{1}|$. Equation (\ref{M36}) is
clearly the same as 
\begin{equation}
\langle V_{1 ||}\rangle=\biggl \langle {{\bf V}_{0}{\bf V}_{1}\over { V}_{0}}\biggr \rangle= \int W({\bf V}_{0},{\bf V}_{1}) {\bf V}_{1}
{\mb 1}_{0} d^{2}{\bf V}_{0}d^{2}{\bf V}_{1}
\label{M37}
\end{equation} 
Substituting for $\langle V_{1 ||}\rangle_{{ V}_{0}}$ from equation (\ref{M27})
in equation (\ref{M36}), we have
\begin{equation}
\langle V_{1 ||}\rangle=-{ n\gamma^{2}\over 2\pi}\int_{0}^{+\infty} V_{0}
dV_{0}\int_{0}^{+\infty}\rho^{2}d\rho e^{-n C(\rho)}J_{1}(\rho
V_{0})Q(\rho,r_{1})
\label{M38}
\end{equation} 
It is apparent that we cannot interchange the order of integration in equation
(\ref{M38}). Following a method similar to that adopted in section
\ref{sec_V1moy}, we rewrite equation (\ref{M38}) in the form:
\begin{equation}
\langle V_{1 ||}\rangle=\int_{0}^{+\infty} dV_{0}\int_{0}^{+\infty}
V_{0}J_{1}(\rho V_{0})\Phi(\rho)d\rho
\label{M39}
\end{equation} 
where
\begin{equation}
\Phi(\rho)=-{n\gamma^{2}\over 2\pi}\rho^{2} e^{-n C(\rho)}   Q(\rho,r_{1})
\label{Phirho}
\end{equation}
Integrating by parts the second integral with the identity (\ref{id15})
we obtain
\begin{equation}
\langle V_{1 ||}\rangle=\int_{0}^{+\infty} dV_{0} \int_{0}^{+\infty} J_{0}(\rho
V_{0})\Phi'(\rho)d\rho
\label{M40}
\end{equation} 
Under this form, it is now possible to interchange the order of integration.
Using formula (\ref{id9}), we find that
\begin{equation}
\langle V_{1 ||}\rangle=\int_{0}^{+\infty} 
 {\Phi'(\rho)\over\rho} d\rho
\label{M41}
\end{equation}
Again integrating by parts we obtain
\begin{equation}
\langle V_{1 ||}\rangle=\int_{0}^{+\infty} 
 {\Phi(\rho)\over\rho^{2}} d\rho=-{n\gamma^{2}\over 2\pi} \int_{0}^{+\infty} 
e^{-n C(\rho)}Q(\rho,r_{1})d\rho
\label{M42}
\end{equation}

In Appendix \ref{sec_M42}, we derive an alternative form of equation 
(\ref{M42}) in terms of $\Gamma$-functions. Indeed, we show that
\begin{equation}
\langle V_{1 ||}\rangle=\langle |{\bf V}_{0}|\rangle-\biggl (
{n\gamma^{2}\over\pi \ln N}\biggr )^{1/2}\int_{0}^{+\infty}dz {J_{1}(z)\over
z^{2}}\int_{0}^{s^{2}z^{2}}e^{-t}t^{-1/2}dt
\label{M49}
\end{equation} 
where we have introduced the notation 
\begin{equation}
s=\biggl ( {\pi n\ln N\over 4}\biggr )^{1/2}r_{1}
\label{sbis}
\end{equation}
to measure the distance $r_{1}$ in terms of the average distance between
vortices. Defining the incomplete $\Gamma$-function by:
\begin{equation}
\Gamma_{x}(p+1)=\int_{0}^{x}e^{-t}t^{p}dt
\label{gamma}
\end{equation} 
we have
\begin{equation}
\langle V_{1 ||}\rangle=\langle |{\bf V}_{0}|\rangle-\biggl (
{n\gamma^{2}\over\pi \ln N}\biggr )^{1/2}\int_{0}^{+\infty}{J_{1}(z)\over
z^{2}}\Gamma_{s^{2}z^{2}}\biggl ({1\over 2}\biggr )dz
\label{M50}
\end{equation} 
This result can be compared with formula (117) of Chandrasekhar \cite{c3}
for the correlation in the gravitational forces acting at two points separated
by a finite distance.

The dependence upon $r_{1}$, through the variable $s$ is encapsulated in the
function
\begin{equation}
{f}(s)=\int_{0}^{+\infty}{J_{1}(z)\over z^{2}}\Gamma_{s^{2}z^{2}}\biggl ({1\over
2}\biggr )dz
\label{f1}
\end{equation} 
In Appendix \ref{sec_f}, it is  shown that
\begin{equation}
{f}(s)\simeq 2 s \qquad (s\rightarrow 0)
\label{f2}
\end{equation} 
\begin{equation}
{f}(s)\simeq 1.41548...+{\sqrt{\pi}\over 2}\ln s  \qquad (s\rightarrow +\infty)
\label{f3}
\end{equation}
This leads to the asymptotic behaviours
\begin{equation}
\langle V_{1 ||}\rangle=\langle |{\bf V}_{0}|\rangle-n\gamma r_{1}+... \qquad
(r_{1}\rightarrow 0)
\label{M51}
\end{equation} 
\begin{equation}
\langle V_{1 ||}\rangle={2\over \ln N}\langle |{\bf V}_{0}|\rangle  \ln\biggl
({R\over r_{1}}\biggr ) \qquad (r_{1}\rightarrow +\infty)
\label{M52}
\end{equation} 
consistent with formulae (\ref{M31}) and  (\ref{M34})-(\ref{M35}). We
observe that the velocity correlations decay extremely slowly 
with the distance $r_{1}$ while the initial slope of the function $\langle V_{1
||}\rangle$ goes rapidly to zero at $d\sim n^{-1/2}$, the interparticle
distance (see equation (\ref{M51})).

\subsection{The function $\langle {{\bf V}_{0}{\bf V}_{1}\over V_{0}^{2}}\rangle$}
\label{sec_foncnorm}

We now wish  to evaluate the function
\begin{equation}
K(r_{1})=\biggl\langle {{\bf V}_{0}{\bf V}_{1}\over V_{0}^{2}}\biggr\rangle
\label{norm1bis}
\end{equation}  
According to equation (\ref{M38}), we immediately have
\begin{equation}
K(r_{1})=-{ n\gamma^{2}\over 2\pi}\int_{0}^{+\infty}
dV_{0}\int_{0}^{+\infty}\rho^{2}d\rho e^{-n C(\rho)}J_{1}(\rho
V_{0})Q(\rho,r_{1})
\label{norm2}
\end{equation} 
Interchanging the order of integrations and using the identity (\ref{id9}), we
obtain 
\begin{equation}
K(r_{1})=-{ n\gamma^{2}\over 2\pi}\int_{0}^{+\infty}\rho d\rho e^{-n
C(\rho)}Q(\rho,r_{1})
\label{norm3}
\end{equation} 
This function is related to equation (\ref{M42}) simply by introducing $\rho$ in the
integral. We can therefore repeat the steps leading to equation (\ref{M50}) and
we obtain instead:
\begin{equation}
K(s)=1-{4\over\ln N}\int_{0}^{+\infty}{J_{1}(z)\over
z^{2}}\Gamma_{s^{2}z^{2}}(1)dz
\label{norm4}
\end{equation} 
where $s$ is defined by equation (\ref{sbis}). Explicitly, it has the form: 
\begin{equation}
K(s)=1-{4\over\ln N}\int_{0}^{+\infty}{J_{1}(z)\over z^{2}}
(1-e^{-s^{2}z^{2}})  dz
\label{norm5}
\end{equation} 
It turns out that this integral can be expressed in terms of known functions.
Indeed:
\begin{equation}
K(s)=1-{4s^{2}\over\ln N}\biggl (1-e^{-{1\over 4 s^{2}}}+{1\over 4
s^{2}}E_{1}\biggl ({1\over 4 s^{2}}\biggr )\biggr )
\label{norm6}
\end{equation} 
where $E_{1}(z)$ denotes the exponential integral
\begin{equation}
E_{1}(z)=\int_{z}^{+\infty}{e^{-t}\over t}dt
\label{norm7}
\end{equation} 
It has the series expansion 
\begin{equation}
E_{1}(z)=C-\ln z-\sum_{n=1}^{+\infty}{(-1)^{n}z^{n}\over n!n}
\label{norm8}
\end{equation} 
where $C=0.5772...$ is the Euler constant. For large values of the argument
$z\rightarrow +\infty$, we have:
\begin{equation}
E_{1}(z)\sim {e^{-z}\over z}\biggl (1-{1\over z}+{2\over z^{2}}-{3!\over
z^{3}}+...\biggr )
\label{norm9}
\end{equation} 
Using these formulae, we obtain the following behaviour of $K(r_{1})$ for
small and large separations: 
\begin{equation}
K(r_{1})=1-\pi n r_{1}^{2}+...\qquad (r_{1}\rightarrow 0)
\label{norm10}
\end{equation} 
\begin{equation}
K(r_{1})= {2\over \ln N}\ln\biggl ({R\over r_{1}}\biggr)\qquad
(r_{1}\rightarrow +\infty)
\label{norm11}
\end{equation}

\section{The spatial velocity correlation function and the energy spectrum}
\label{sec_corrfonction}

\subsection{The  correlation function $\langle {\bf V}_{0}{\bf V}_{1}\rangle$}
\label{sec_fonc}

The spatial correlation function of the velocity is defined by:
\begin{equation}
\langle {\bf V}_{0}{\bf V}_{1}\rangle=\int W({\bf V}_{0},{\bf V}_{1}){\bf
V}_{0}{\bf V}_{1}d^{2}{\bf V}_{0}d^{2}{\bf V}_{1}
\label{N1}
\end{equation}
It is related to equation (\ref{M37}) by simply introducing $V_{0}$ in the
integral. Therefore, according to equation (\ref{M40}), we have immediately:
\begin{equation}
\langle {\bf V}_{0}{\bf V}_{1}\rangle=\int_{0}^{+\infty}dV_{0}
\int_{0}^{+\infty} \rho V_{0}J_{0}(\rho V_{0}){\Phi'(\rho)\over\rho} d\rho
\label{N2}
\end{equation}
where $\Phi(\rho)$ is defined by equation (\ref{Phirho}). Using the identity
(\ref{id13}) and integrating by parts the second integral, we obtain:
\begin{equation}
\langle {\bf V}_{0}{\bf
V}_{1}\rangle=-\int_{0}^{+\infty}dV_{0}\int_{0}^{+\infty}\rho J_{1}(\rho
V_{0}){d\over d\rho}\biggl( {\Phi '(\rho)\over\rho}\biggr )d\rho
\label{N3}
\end{equation}
Inverting the order of integration and using (\ref{id9}), we find
\begin{equation}
\langle {\bf V}_{0}{\bf V}_{1}\rangle=-\int_{0}^{+\infty} {d\over d\rho}\biggl
({\Phi ' (\rho)\over\rho}\biggr )d\rho
\label{N4}
\end{equation}
The correlation function (\ref{N1}) therefore takes the relatively 
simple form:  
\begin{equation}
\langle {\bf V}_{0}{\bf V}_{1}\rangle=\lim_{\rho\rightarrow 0}{\Phi
'(\rho)\over\rho}
\label{N5}
\end{equation}
Using equations (\ref{Phirho}) and (\ref{C2}), the derivative of $\Phi$ is
\begin{eqnarray}
\Phi '(\rho)={n^{2}\gamma^{4}\over 16\pi^{2}}\rho^{3}\ln\biggl ({4\pi N\over
n\gamma^{2}\rho^{2}}\biggr ) e^{-nC(\rho)}Q(\rho, r_{1})\nonumber\\
-{n\gamma^{2}\over\pi}\rho   e^{-nC(\rho)}    Q(\rho, r_{1})- {n\gamma^{2}\over
2\pi}   e^{-nC(\rho)}     Q'(\rho, r_{1})\rho^{2}
\label{Phiprime}
\end{eqnarray}
For $\rho\rightarrow 0$ equation (\ref{Q4}) reduces to:
\begin{equation}
Q(\rho,r_{1})=- {1\over 2}\ln\biggl ({R\over r_{1}}\biggr )+O(\rho^{2})
\label{Qapp}
\end{equation}
Therefore, according to equation (\ref{N5}), we obtain: 
\begin{equation}
\langle {\bf V}_{0}{\bf V}_{1}\rangle={n\gamma^{2}\over 2\pi}\ln\biggl ({R\over
r_{1}}\biggr )
\label{N6}
\end{equation}
Like the  quantities (\ref{M37}) (\ref{norm1bis}), the velocity 
autocorrelation 
function (\ref{N6}) decays extremely slowly at large distances. However, unlike (\ref{M37}) and (\ref{norm1bis}), the velocity autocorrelation
function diverges logarithmically at small separations since the variance
$\langle V^{2}\rangle$ does not exist (see section \ref{sec_vpdf}). In Appendix \ref{sec_alt}, we give a more direct derivation of formula (\ref{N6}).

\subsection{The energy spectrum of a random distribution of point vortices}
\label{sec_spectrum}

The correlation function (\ref{N1}) is related to the energy spectrum of the
random distribution of point vortices by the Fourier transform
\begin{equation}
\langle {\bf V}(0){\bf V}({\bf r})\rangle={1\over 4\pi^{2}}\int e^{-i{\bf k}{\bf
r}}{4\pi E(k)\over k}d^{2}{\bf k}
\label{S1}
\end{equation}
Inversely,
\begin{equation}
{4\pi E(k)\over k}=\int e^{i{\bf k}{\bf r}}\langle {\bf V}(0){\bf V}({\bf
r})\rangle d^{2}{\bf r}
\label{S2}
\end{equation}
Introducing polar coordinates and carrying out the integration  over the angular
variable, we find
\begin{equation}
E(k)={1\over 2}\int_{0}^{+\infty} J_{0}(kr) \langle {\bf V}(0){\bf V}({\bf
r})\rangle k r dr
\label{S3}
\end{equation}
According to equation (\ref{N6}), we have for $r\le R$:
\begin{equation}
\langle {\bf V}(0){\bf V}({r})\rangle={n\gamma^{2}\over 2\pi}\ln\biggl ({R\over
r}\biggr ) 
\label{S4}
\end{equation}
Hence
\begin{equation}
E(k)={n\gamma^{2}\over 4\pi}\int_{0}^{R} J_{0}(kr) \ln\biggl ({R\over r}\biggr )
k r dr
\label{S5}
\end{equation}
Using (\ref{id13}) and integrating by parts, we get
\begin{equation}
E(k)={n\gamma^{2}\over 4\pi}\int_{0}^{R} J_{1}(kr) dr
\label{S6}
\end{equation}
Again integrating by parts  with the identity (\ref{id15}), we obtain
\begin{equation}
E(k)={n\gamma^{2}\over 4\pi k}(1-J_{0}(kR))
\label{S7}
\end{equation} 
For large $k$ the energy spectrum reduces to
\begin{equation}
E(k)\sim{n\gamma^{2}\over 4\pi k} \qquad (k\rightarrow +\infty)
\label{S8}
\end{equation}
which is Novikov \cite{novikov} result. At small $k$, using
equation (\ref{S7}), we find
\begin{equation}
E(k)\sim {N\gamma^{2}\over 16\pi^{2}}k \qquad (k\rightarrow 0)
\label{S9}
\end{equation}

\section{Conclusion}
\label{sec_conclusion}

In this paper, we have analyzed in some details the statistical features of the stochastic velocity field produced by a random distribution of point vortices in two dimensions. In particular, we have obtained exact results characterizing the correlations in the velocities occuring at two points separated by an arbitrary distance. We have derived an explicit expression for the spatial velocity autocorrelation function and found that the correlations decay extremely slowly with the distance. The other quantities computed in this article are less standard quantities in turbulence  but we do not see any reason why they should not be considered in details. They could  be measured in simulations of point vortex dynamics and this would provide a direct confrontation between our theoretical model and a situation in which the temporal dynamics of the point vortices is explicitly taken into account. One interest of our model is to provide exact results for the velocity correlations,  which is not so frequent in turbulence.  

Our calculations could be relevant to the context of decaying two dimensional turbulence when the flow becomes dominated by a large number of coherent vortices. Indeed, our model is based on the same assumptions as in Ref. \cite{min,jimenez,weiss} and these assumptions have been vindicated by Direct Navier Stokes simulations and laboratory experiments. In particular, the assumption that the Poisson distribution is stationary is vindicated both by theoretical arguments and by the simulations of Jim\'enez. Of course, in decaying turbulence the density of vortices varies with time but it should be possible to integrate this dependance in the theory to make our results useful in more general situations. Also, in two dimensional turbulence the vortices have a finite core (vortex ``blobs'') and this can severely alter the predictions of the point vortex model. However, our formalism is general and we can introduce a lower cut off in the 
theory to take into account finite size effects. For the first time, we have obtained explicit expressions for the distribution of velocity and 
velocity increments  taking into account the finite size of 
the vortices. For 
``extended'' vortices we have proved that the p.d.f. of both velocity and velocity differences are Gaussian. Extended vortices occur in the early stage of 2D decaying turbulence (when the area covered by the vortices is still large \cite{jimenez}) or when the Reynolds number is low \cite{bracco}. Gaussian p.d.f. are indeed observed in these situations and have been discussed in Ref. \cite{jimenez,bracco}. At the late stages of the decline (when the area covered by the vortices has decreased \cite{jimenez}) or for high Reynolds numbers \cite{bracco} the point vortex model should be more and more accurate. It would predict a marginal Gaussian distribution (with an algebraic tail) for the velocity p.d.f. and a Cauchy law for the velocity differences. However, our analytic results show that for ``small'' but non singular vortices the p.d.f can substantially deviate from the case $a= 0$ and this can possibly explain (in a quasi-analytical framework) the occurence of exponential tails observed and discussed in Ref. \cite{min,jimenez,bracco} (this point will be developed elsewhere). 

In the present study, we have exclusively considered the {\it spatial} correlations of the velocity occuring between two points at the same time.  Another quantity of fundamental interest is the {\it temporal} correlation function whose integral 
determines the diffusion coefficient of point vortices through a Kubo formula 
\cite{chavkin}. It is not possible, however, to relate the spatial and 
temporal correlations functions by considering that the point vortices 
follow linear trajectories with uniform velocity as is commonly done in 
the case of stars or electric charges \cite{c4}. Physically, the linear 
trajectory approximation made in plasma physics or in stellar dynamics 
is not applicable here because the vortices, unlike material particles, 
do not have inertia so they do not move on their own. There is, however, a 
situation where such an assumption can be implemented. It concerns the motion 
of point vortices in the presence of a strong background shear as 
investigated by Chavanis \cite{chav98a,chavkin}. In that situation we 
can consider, 
to a first approximation, that the point vortices follow the streamlines 
of the shear and evaluate the temporal correlation function within this 
assumption. The correlation function is found to decay as $t^{-2}$ for 
$t\rightarrow +\infty$. This is a slow decay but sufficient to insure the 
convergence of the diffusion coefficient. In addition, when the vortices 
move in a background shear they experience a {\it systematic drift} 
\cite{chav98a} normal 
to their mean field velocity. The drift coefficient is proportional to 
the diffusion coefficient (hence to the velocity correlation function) through 
an Einstein relation, like in Brownian theory. At equilibrium the drift balances the scattering 
and maintains an inhomogeneous vortex distribution. This drift 
is of great importance to understand the organization of point vortices 
at negative temperatures \cite{onsager} and has been discussed in details
 elsewhere 
\cite{chav98a,chavkin}. In the absence of background shear, we cannot evaluate the temporal correlation function simply but we can estimate the diffusion coefficient of point vortices through the analysis of the velocity fluctuations performed by Chavanis \& Sire \cite{cs}. Therefore, these results are of great importance to build up a rational kinetic theory of point vortices \cite{chav98a,chavkin}.

\acknowledgements 

One of us (PHC) would like to acknowledge interesting discussions with I. Mezic during the program on Hydrodynamic Turbulence at the Institute of Theoretical Physics, Santa Barbara.

\newpage
\appendix

\section{Detailed calculations of subsection III D}
\label{sec_detailed1}

In this Appendix, we determine the asymptotic behaviour  for $|{\mb \sigma}| \rightarrow 0$ of the function $A({\mb
\rho},{\mb \sigma})$ defined in subsection \ref{sec_deltaV}. According to equation (\ref{A7}), we need the behaviour of $C({\mb
\rho},{\mb \sigma})$ for $|{\mb \sigma}| \rightarrow 0$.  For ${\mb \sigma}={\mb
0}$, $C({\mb \rho},{\mb \sigma})$ reduces to the function $C({\mb \rho})$
introduced in section \ref{sec_vpdf}. Writing
\begin{equation}
C({\mb \rho},{\mb \sigma})=C({\mb\rho})+D({\mb \rho},{\mb \sigma})
\label{C10}
\end{equation}
we have
\begin{equation}
D({\mb \rho},{\mb \sigma})={\gamma^{2}\over 4\pi^{2}}\int_{|{\bf
\Phi}|={0}}^{+\infty}     e^{i{\mb \rho}{\mb \Phi}}(1-e^{i{\mb \sigma}{\mb
\psi}}){1\over \Phi^{4}}d^{2}{\bf \Phi}
\label{D1}
\end{equation}
We have let $R\rightarrow +\infty$ since the integral (\ref{D1}) is convergent
when $|{\mb\Phi}|\rightarrow 0$. For $|{\mb \sigma}| \rightarrow 0$, we can
expand the exponential term $e^{i{\mb \sigma}{\mb \psi}}$ which occurs under the
integral sign in equation (\ref{D1}) in a power series in ${\mb \sigma}$.
Retaining only the first term in this expansion, we have
\begin{equation}
D({\mb \rho},{\mb \sigma})=-{\gamma^{2}\over 4\pi^{2}}i\int_{|{\bf
\Phi}|=0}^{+\infty} e^{i{\mb \rho}{\mb \Phi}}({\mb \sigma} {\mb \psi}){1\over
\Phi^{4}}d^{2}{\bf \Phi}+o(|{\mb\sigma}|^{2}) 
\label{D2}
\end{equation}
Substituting for ${\mb \psi}$ from equation (\ref{Psi3}) in equation (\ref{D2}),
we obtain 
\begin{equation}
D({\mb \rho},{\mb \sigma})=i{\gamma\over 2\pi}\int_{|{\bf \Phi}|=0}^{+\infty}
\cos({\mb \rho}{\mb \Phi})\biggl \lbrace \Phi^{2}{\mb \sigma}\delta {\bf
r}_{\perp}+ 2 ({\mb \Phi}_{\perp}\delta {\bf r})({\mb \sigma}{\mb \Phi})\biggr
\rbrace {d^{2}{\mb \Phi}\over \Phi^{4}}+o(|{\mb\sigma}|^{2}) 
\label{D3}
\end{equation}
To evaluate this integral, we introduce a Cartesian system of coordinates where
the $x$-axis is in the direction of ${\mb\rho}$. We denote by $({\delta
r}_{1},{\delta  r}_{2})$ and $(\sigma_{1},\sigma_{2})$ the components of $\delta
{\bf r}$ and ${\mb \sigma}$ in this system of coordinates and we introduce
$\theta$, the angle that ${\mb\Phi}$ forms with ${\mb\rho}$. Transforming to
polar coordinates and setting $x=\rho\Phi$, we obtain after some rearrangements:
\begin{eqnarray}
D({\mb \rho},{\mb \sigma})=i{\gamma\over
2\pi}\int_{0}^{2\pi}d\theta\int_{0}^{+\infty}{dx\over x}
\cos(x\cos\theta)\nonumber\\
\times\biggl \lbrace (\sigma_{1}\delta r_{2}+\sigma_{2}\delta
r_{1})\cos(2\theta)+(\sigma_{2}\delta r_{2}-\sigma_{1}\delta r_{1})\sin
(2\theta)\biggr \rbrace +o(|{\mb\sigma}|^{2})  
\label{D5}
\end{eqnarray}
Using the expansion
\begin{equation}
\cos (x\cos\theta)=J_{0}(x)+2\sum_{n=1}^{+\infty}(-1)^{n}J_{2n}(x)\cos
(2n\theta)
\label{id2bis}
\end{equation}
and the identities
\begin{equation}
\int_{0}^{2\pi}\cos(m\theta)\cos(n\theta)d\theta=\pi \delta_{nm}
\label{id3}
\end{equation}
\begin{equation}
\int_{0}^{2\pi}\sin(m\theta)\sin(n\theta)d\theta=\pi\delta_{nm}
\label{id5}
\end{equation}
\begin{equation}
\int_{0}^{2\pi}\cos(m\theta)\sin(n\theta)d\theta=0
\label{id6}
\end{equation}
we find that
\begin{equation}
D({\mb \rho},{\mb \sigma})=-i\gamma (\sigma_{1}\delta r_{2}+\sigma_{2}\delta
r_{1}) \int_{0}^{+\infty} {J_{2}(x)\over x}dx+o(|{\mb\sigma}|^{2}) 
\label{D6}
\end{equation}
Since
\begin{equation}
\int_{0}^{+\infty}J_{2}(x){dx\over x}={1\over 2}
\label{id4}
\end{equation}
we finally obtain
\begin{equation}
D({\mb \rho},{\mb \sigma})=-i{{\gamma}\over 2} (\sigma_{2}{\delta
r}_{1}+\sigma_{1}{\delta r}_{2})+o(|{\mb\sigma}|^{2}) 
\label{D7}
\end{equation} 
Now, according to (\ref{A7})(\ref{C10}) and (\ref{D7}), we have:
\begin{equation}
A({\mb\rho},{\mb\sigma})= e^{-n C({\mb\rho})+i{{\gamma} n\over
2}(\sigma_{2}\delta r_{1}+\sigma_{1}\delta r_{2})+o(|{\mb\sigma}|^{2})} 
\label{A10}
\end{equation} 
where we recall that $({\delta  r}_{1},{\delta  r}_{2})$ and
$(\sigma_{1},\sigma_{2})$ are the components of $\delta {\bf r}$ and ${\mb
\sigma}$ in a system of coordinates where the $x$-axis is in the direction of
${\mb\rho}$. Now, to evaluate the integral (\ref{dVV5}) we must express $\delta
{\bf r}$ and ${\mb \sigma}$ in a fixed system of coordinates independent on
${\mb\rho}$. We choose this system such that the $x$-axis coincides with the
direction of $\delta {\bf r}$. If $\theta$ denotes the angle that ${\mb\rho}$
forms with  $\delta {\bf r}$, the components $(\sigma_{x},\sigma_{y})$ of
${\mb\sigma}$ in this system are related to $(\sigma_{1},\sigma_{2})$ by the
transformations:  
\begin{equation}
\sigma_{1}=\sigma_{x}\cos\theta+\sigma_{y}\sin\theta
\label{T1}
\end{equation}    
\begin{equation}
\sigma_{2}=\sigma_{y}\cos\theta-\sigma_{x}\sin\theta
\label{T2}
\end{equation} 
Using in addition $\delta r_{1}=|\delta{\bf r}|\cos\theta$ and $\delta
r_{2}=-|\delta{\bf r}|\sin\theta$,  $A({\mb\rho},{\mb\sigma})$ can be rewritten
in the form (\ref{A11}).

\section{Detailed calculations of subsection IV A}
\label{sec_detailed2}

In this Appendix, we determine the expression of the function  $D^{(1)}({\mb \rho},{\mb \sigma})$ defined by equation (\ref{D10}). This function can be written alternatively:
\begin{equation}
D^{(1)}({\mb \rho},{\mb \sigma})=-{i\gamma\over 2\pi}\int_{|{\bf r}|=0}^{R}
e^{i{\gamma\over 2\pi}{{\mb \rho}_{\perp}{\bf r}\over r^{2}}}{{\mb
\sigma}_{\perp}}{\bf \nabla}(\ln |{\bf r}-{\bf r}_{1}|) d^{2}{\bf r}
\label{D11}
\end{equation} 
Integrating by parts, we find
\begin{eqnarray}
D^{(1)}({\mb \rho},{\mb \sigma})=-{i\gamma\over 2\pi}\int_{|{\bf
r}|=0}^{R}\nabla \biggl \lbrace e^{i{\gamma\over 2\pi}{{\mb \rho}_{\perp}{\bf
r}\over r^{2}}}\ln |{\bf r}-{\bf r}_{1}|{\mb \sigma}_{\perp} \biggr \rbrace
d^{2}{\bf r} \nonumber\\  +{i\gamma\over 2\pi} \int_{|{\bf r}|=0}^{R}\ln |{\bf
r}-{\bf r}_{1}| {\mb \sigma}_{\perp}    \nabla \biggl ( e^{i{\gamma\over
2\pi}{{\mb \rho}_{\perp}{\bf r}\over r^{2}}}\biggr ) d^{2}{\bf r}
\label{D12}
\end{eqnarray} 
Explicitly, it has the form
\begin{eqnarray}
D^{(1)}({\mb \rho},{\mb \sigma})=-{i\gamma\over 2\pi}   \oint_{|{\bf r}|=R}
e^{i{\gamma\over 2\pi}{{\mb \rho}_{\perp}{\bf r}\over r^{2}}}\ln |{\bf r}-{\bf
r}_{1}| {\mb \sigma}_{\perp}     dl \hat{{\bf e}}_{r}\nonumber\\
-{\gamma^{2}\over 4\pi^{2}}\int_{|{\bf r}|=0}^{R} e^{i{\gamma\over 2\pi}{{\mb
\rho}_{\perp}{\bf r}\over r^{2}}}\ln |{\bf r}-{\bf r}_{1}|\biggl \lbrace {{\mb
\rho}_{\perp} {\mb \sigma}_{\perp}  \over r^{2}}-{2 ({\mb \rho}_{\perp}   {\bf
r})( {\mb \sigma}_{\perp}    {\bf r})\over r^{4}}\biggr\rbrace d^{2}{\bf r}
\label{D13}
\end{eqnarray} 
The first integral is a boundary term which behaves like $\ln R$ when
$R\rightarrow\infty$; we call it:
\begin{equation}
D^{(1)}_{boundary}({\mb \rho},{\mb \sigma})=-{i\gamma\over 2\pi}   \oint_{|{\bf
r}|=R}  e^{i{\gamma\over 2\pi}{{\mb \rho}_{\perp}{\bf r}\over r^{2}}}\ln |{\bf
r}-{\bf r}_{1}| {\mb \sigma}_{\perp}     dl \hat{{\bf e}}_{r}
\label{D14}
\end{equation} 
The second integral is convergent when $R\rightarrow\infty$; we call it:
\begin{equation}
D^{(1)}_{int}({\mb \rho},{\mb \sigma}) =-{\gamma^{2}\over 4\pi^{2}}\int_{|{\bf
r}|=0}^{+\infty} e^{i{\gamma\over 2\pi}{{\mb \rho}_{\perp}{\bf r}\over
r^{2}}}\ln |{\bf r}-{\bf r}_{1}|\biggl \lbrace {{\mb \rho}_{\perp} {\mb
\sigma}_{\perp}  \over r^{2}}-{2 ({\mb \rho}_{\perp}   {\bf r})( {\mb
\sigma}_{\perp}    {\bf r})\over r^{4}}\biggr\rbrace d^{2}{\bf r}
\label{D16}
\end{equation} 
Then
\begin{equation}
D^{(1)}({\mb \rho},{\mb \sigma})=D^{(1)}_{boundary}({\mb \rho},{\mb
\sigma})+D^{(1)}_{int}({\mb \rho},{\mb \sigma})
\label{D15}
\end{equation}

\subsection{The evaluation of $D^{(1)}_{int}({\mb \rho},{\mb \sigma})$}
\label{sec_Dint}

To evaluate $D^{(1)}_{int}({\mb \rho},{\mb \sigma})$, we introduce a Cartesian
system of coordinates with the $x$-axis is the direction of ${\mb
\rho}_{\perp}$. We denote by $(\sigma_{1},\sigma_{2})$ the components of
${\mb\sigma}$ in this system. Introducing $\theta$, the angle that ${\bf r}$
forms with ${\mb \rho}_{\perp}$ and transforming to polar coordinates, we
obtain:
\begin{equation}
D^{(1)}_{int}({\mb \rho},{\mb \sigma})= {\gamma^{2}\over
4\pi^{2}}\rho\int_{0}^{2\pi}d\theta\int_{0}^{+\infty}{dr\over r}
e^{i{\gamma\rho\cos\theta\over 2\pi r}}\ln |{\bf r}-{\bf r}_{1}|\lbrace
\sigma_{2}+2\cos\theta (\sigma_{1}\sin\theta -\sigma_{2}\cos\theta)\rbrace
\label{D17}
\end{equation}     
or, equivalently,
\begin{equation}
D^{(1)}_{int}({\mb \rho},{\mb \sigma})= {\gamma^{2}\over 4\pi^{2}}\rho
\int_{0}^{2\pi}d\theta\int_{0}^{+\infty}{dr\over r}
e^{i{\gamma\rho\cos\theta\over 2\pi r}}\ln |{\bf r}-{\bf r}_{1}|\lbrace
\sigma_{1}\sin(2\theta) -\sigma_{2}\cos(2\theta)\rbrace
\label{D18}
\end{equation}
We now expand $e^{i{\gamma\rho\cos\theta\over 2\pi r}}$ and $\ln |{\bf r}-{\bf
r}_{1}|$ which occur under the integral sign in terms of sinusoidal functions
using the identities:
\begin{equation}
e^{ix\cos\theta}=J_{0}(x)+2\sum_{n=1}^{+\infty}i^{n}J_{n}(x)\cos(n\theta)
\label{id10}
\end{equation}     
\begin{equation}
\ln |{\bf r}-{\bf r}_{1}|=\ln r_{>}-\sum_{m=1}^{+\infty}{1\over m}\biggl
({r_{<}\over r_{>}}\biggr )^{m}\cos\lbrack m(\theta-\theta_{1})\rbrack
\label{id11}
\end{equation} 
where $J_{n}(x)$ is the Bessel function of order $n$,  $r_{>}$ (resp. $r_{<}$)
is the larger (resp. smaller) of $(r,r_{1})$ and $\theta_{1}$ is the angle  that
${\bf r}_{1}$ forms with ${\mb\rho}_{\perp}$.

Using the orthogonality properties (\ref{id3}) (\ref{id5}) (\ref{id6}), we can
express $D^{(1)}_{int}({\mb \rho},{\mb \sigma})$ in the form:
\begin{eqnarray}
D^{(1)}_{int}({\mb \rho},{\mb \sigma})={\gamma^{2}\over
2\pi}\rho\sigma_{2}\int_{0}^{+\infty}{dr\over r}J_{2}\biggl ( {\gamma\rho\over
2\pi r}\biggr )\ln r_{>}\nonumber\\
-\sigma_{2}\sum_{n=1}^{+\infty}A_{n}(\rho,r_{1})\cos
(n\theta_{1})-\sigma_{1}\sum_{n=1}^{+\infty}B_{n}(\rho,r_{1})\sin (n\theta_{1})
\label{D19}
\end{eqnarray}
where:
\begin{equation}
A_{n}(\rho,r_{1})={\gamma^{2}\over 4\pi}\rho {i^{n}\over
n}\int_{0}^{+\infty}{dr\over r}\biggl ({r_{<}\over r_{>}}\biggr )^{n}\biggl
\lbrack J_{n+2} \biggl ( {\gamma\rho\over 2\pi r}\biggr )+J_{n-2} \biggl (
{\gamma\rho\over 2\pi r}\biggr )\biggr\rbrack
\label{An1}
\end{equation}
\begin{equation}
B_{n}(\rho,r_{1})={\gamma^{2}\over 4\pi}\rho {i^{n}\over
n}\int_{0}^{+\infty}{dr\over r}\biggl ({r_{<}\over r_{>}}\biggr )^{n}\biggl
\lbrack J_{n+2} \biggl ( {\gamma\rho\over 2\pi r}\biggr )-J_{n-2} \biggl (
{\gamma\rho\over 2\pi r}\biggr )\biggr\rbrack
\label{Bn1}
\end{equation}

\subsection{The evaluation of $D^{(1)}_{boundary}({\mb \rho},{\mb \sigma})$}
\label{sec_Dboundary}

Using the same system of coordinates, we can rewrite our expression for
$D^{(1)}_{boundary}({\mb \rho},{\mb \sigma})$  in the form:
\begin{equation}
D^{(1)}_{boundary}({\mb \rho},{\mb \sigma})=-i {\gamma\over 2\pi}
R\int_{0}^{2\pi}d\theta  e^{i{\gamma\rho\cos\theta\over 2\pi r}}\ln |{\bf
r}-{\bf r}_{1}|(\sigma_{1}\sin\theta-\sigma_{2}\cos\theta)
\label{D20}
\end{equation}
Substituting for (\ref{id10}) and (\ref{id11}) in equation (\ref{D20}) and
carrying out the angular integration, we obtain
\begin{eqnarray}
D^{(1)}_{boundary}({\mb \rho},{\mb \sigma})=-\gamma  R \sigma_{2} J_{1}\biggl
({\gamma\rho\over 2\pi R  }\biggr )\ln R \nonumber\\ -{1\over 2}\gamma
R\sigma_{2}  \sum_{n=1}^{+\infty}{i^{n}\over n}\biggl ({r_{1}\over R}\biggr
)^{n}\biggl \lbrack J_{n-1} \biggl ( {\gamma\rho\over 2\pi R}\biggr )-J_{n+1}
\biggl ( {\gamma\rho\over 2\pi R  }\biggr )\biggr\rbrack\cos
(n\theta_{1})\nonumber\\ +{1\over 2}\gamma R\sigma_{1}
\sum_{n=1}^{+\infty}{i^{n}\over n}\biggl ({r_{1}\over R}\biggr )^{n}\biggl
\lbrack J_{n-1} \biggl ( { {\gamma\rho\over 2\pi R  }   }\biggr )+J_{n+1} \biggl
( {\gamma\rho\over 2\pi R  } \biggr )\biggr\rbrack\sin (n\theta_{1})
\label{D21}
\end{eqnarray}
When  $R\rightarrow\infty$, the function $D^{(1)}_{boundary}({\mb \rho},{\mb
\sigma})$ takes the relatively simple form: 
\begin{equation}
D^{(1)}_{boundary}({\mb \rho},{\mb \sigma})=-{\gamma^{2}\over 4\pi}\rho
\sigma_{2}\ln R+i{\gamma\over 2}
r_{1}(-\sigma_{2}\cos\theta_{1}+\sigma_{1}\sin\theta_{1})
\label{D22}
\end{equation}

\subsection{Final expression of $D^{(1)}({\mb \rho},{\mb \sigma})$}
\label{sec_Alim}

 In
equations (\ref{D19}) and (\ref{D22}), ${\mb\sigma}=(\sigma_{1},\sigma_{2})$ is
refered to a variable system of coordinates depending on the direction of
${\mb\rho}_{\perp}$. Since $W({\bf V}_{0},{\bf V}_{1})$ is the Fourier transform
of $A({\mb\rho},{\mb\sigma})$, related to $D^{(1)}({\mb \rho},{\mb \sigma})$ by equation (\ref{A16}), we need to express the vectors ${\mb\rho}$
and ${\mb\sigma}$ in a fixed system of coordinates. We  choose this system such
that the $x$-axis is in the direction of ${\bf r}_{1}$. The components
$(\sigma_{x},\sigma_{y})$ of $\mb\sigma$ in this system are related to
$(\sigma_{1},\sigma_{2})$ by the transformation: 
\begin{eqnarray}
\sigma_{1}=\sigma_{x}\cos\theta_{1}-\sigma_{y}\sin\theta_{1}
\label{T3}
\end{eqnarray}
\begin{eqnarray}
\sigma_{2}=\sigma_{x}\sin\theta_{1}+\sigma_{y}\cos\theta_{1}
\label{T4}
\end{eqnarray}
where it might be recalled that $-\theta_{1}$ is the angle that
${\mb\rho}_{\perp}$ forms with ${\bf r}_{1}$. Thus, in this new system of
coordinates, $D^{(1)}({\mb \rho},{\mb \sigma})$ has the form (\ref{D23}).

\section{Detailed calculations of subsection VI C}
\label{sec_detailed3}

\subsection{Alternative form of equation (\ref{M42})}
\label{sec_M42}

In this Appendix, we derive an alternative form of equation (\ref{M42}) in terms of $\Gamma$ functions. Substituting for $Q(\rho,r_{1})$ from equation
(\ref{Q4}) in equation (\ref{M42}), we obtain
\begin{equation}
\langle V_{1 ||}\rangle=-{ n\gamma^{2}\over 4\pi}\int_{0}^{+\infty} d\rho e^{-n
C(\rho)}\biggl\lbrace \ln\biggl ({\gamma\rho\over 2\pi R}\biggr
)-\ln\epsilon\biggr\rbrace +I^{(\epsilon)}
\label{M43}
\end{equation}
where
\begin{equation}
I^{(\epsilon)}={ n\gamma^{2}\over 2\pi}\int_{{2\pi r_{1}\over
\gamma}\epsilon}^{+\infty}d\rho e^{-n C(\rho)}\int_{\epsilon}^{{\gamma\rho\over
2\pi r_1}}{J_{1}(z)\over z^{2}}dz
\label{I1}
\end{equation}
and we recall that $\epsilon\rightarrow 0$. Inverting the order of integration, we have
\begin{equation}
I^{(\epsilon)}={ n\gamma^{2}\over 2\pi}\int_{\epsilon}^{+\infty}dz{J_{1}(z)\over
z^{2}} \int_{{2\pi r_{1}\over \gamma}z}^{+\infty}e^{-n C(\rho)}d\rho
\label{I2}
\end{equation}
or, alternatively
\begin{equation}
I^{(\epsilon)}={ n\gamma^{2}\over 2\pi}\int_{\epsilon}^{+\infty}{J_{1}(z)\over
z^{2}}dz \int_{{2\pi r_{1}\over\gamma}\epsilon}^{+\infty} e^{-n C(\rho)}d\rho -
{ n\gamma^{2}\over 2\pi}\int_{\epsilon}^{+\infty}dz {J_{1}(z)\over
z^{2}}\int_{{2\pi r_{1}\over\gamma}\epsilon }^{{2\pi r_{1}\over \gamma}z} e^{-n
C(\rho)}d\rho
\label{I3}
\end{equation}
Using the identity (\ref{id14}), integrating by parts the first integral in
equation (\ref{I3}) and taking the limit $\epsilon\rightarrow 0$, we obtain
\begin{eqnarray}
I^{(\epsilon)}={ n\gamma^{2}\over 2\pi}\biggl\lbrace
\int_{0}^{+\infty}{J_{2}(z)\over z}\ln z dz-{1\over 2}\ln\epsilon\biggr\rbrace
\int_{0}^{+\infty} e^{-n C(\rho)}d\rho\nonumber\\ -{ n\gamma^{2}\over
2\pi}\int_{0}^{+\infty}dz{J_{1}(z)\over z^{2}}\int_{0}^{{2\pi r_{1}\over
\gamma}z} e^{-n C(\rho)}d\rho
\label{I4}
\end{eqnarray}
According to equations (\ref{M43}) and (\ref{I4}) we therefore have
\begin{eqnarray}
\langle V_{1 ||}\rangle=-{ n\gamma^{2}\over 4\pi}\int_{0}^{+\infty}  e^{-n
C(\rho)} \ln\biggl ({\gamma\rho\over 2\pi R}\biggr )d\rho -{ n\gamma^{2}\over
2\pi}\int_{0}^{+\infty}dz{J_{1}(z)\over z^{2}}\int_{0}^{{2\pi r_{1}\over
\gamma}z}d\rho e^{-n C(\rho)}
\label{M44}
\end{eqnarray}
For $r_{1}=0$, there remains
\begin{equation}
\langle V_{0 ||}\rangle=-{ n\gamma^{2}\over 4\pi}\int_{0}^{+\infty} e^{-n
C(\rho)} \ln\biggl ({\gamma\rho\over 2\pi R}\biggr ) d\rho=\biggl
({n\gamma^{2}\over 16}\ln N\biggr )^{1/2}
\label{M45}
\end{equation}
This is precisely the value of  $\langle|{\bf V}_{0}|\rangle$, see equation (\ref{Vmoy}),
as expected from equation (\ref{M37}). Therefore, we can write 
\begin{equation}
\langle V_{1 ||}\rangle=\langle |{\bf V}_{0}|\rangle -{ n\gamma^{2}\over
2\pi}\int_{0}^{+\infty}dz{J_{1}(z)\over z^{2}}\int_{0}^{{2\pi r_{1}\over
\gamma}z} e^{-n C(\rho)}d\rho
\label{M47}
\end{equation}
In the last integral, we can replace the function $C(\rho)$ by its approximate
expression (\ref{C2bis}) without introducing any significant error. Therefore
\begin{equation}
\langle V_{1 ||}\rangle=\langle |{\bf V}_{0}|\rangle -{ n\gamma^{2}\over
2\pi}\int_{0}^{+\infty}dz{J_{1}(z)\over z^{2}}\int_{0}^{{2\pi r_{1}\over
\gamma}z} e^{-{n\gamma^{2}\over 16\pi}\ln N\rho^{2}}d\rho
\label{M48}
\end{equation} 
With the change of variables $t={n\gamma^{2}\over 16\pi}\ln N\rho^{2}$, it takes
the form  (\ref{M49}).

\subsection{The function ${f}(s)$}
\label{sec_f}

The behaviour of $f(s)$ for $s\rightarrow 0$ can be derived from equation
(\ref{f1}) by expanding the incomplete $\Gamma$-function which occurs under the
integral sign as a power series in $s$. To first order in $s$, we have
\begin{eqnarray}
\Gamma_{s^{2}z^{2}}\biggl ({1\over 2}\biggr
)=\int_{0}^{s^{2}z^{2}}e^{-t}t^{-1/2}dt\nonumber\\
=\int_{0}^{s^{2}z^{2}}\biggl (1-t+{t^{2}\over 2}+...\biggr
)t^{-1/2}dt=2 s z +o(s^{3})
\label{f4}
\end{eqnarray} 
Thus 
\begin{equation}
{f}(s)=\int_{0}^{+\infty}{J_{1}(z)\over z^{2}}\Gamma_{s^{2}z^{2}}\biggl ({1\over
2}\biggr )dz=2 s \int_{0}^{+\infty}dz {J_{1}(z)\over z}+...
\label{f5}
\end{equation} 
With the identity (\ref{id2}), we obtain 
\begin{equation}
{f}(s)\simeq 2 s \qquad (s\rightarrow 0)
\label{f6}
\end{equation}

To obtain the behaviour of $f(s)$ for $s\rightarrow +\infty$, we write equation
(\ref{f1}) in the form
\begin{equation}
{f}(s)=\int_{0}^{+\infty}dz {J_{1}(z)\over z^{2}}
\int_{0}^{s^{2}z^{2}}e^{-t}t^{-1/2}dt
\label{f7}
\end{equation} 
and take the derivative with respect to $s$:
\begin{equation}
{f}'(s)=2\int_{0}^{+\infty} {J_{1}(z)\over z} e^{-s^{2}z^{2}}dz
\label{f8}
\end{equation} 
Replacing $J_{1}(z)$ in the foregoing equation by its series expansion
\begin{equation}
J_{1}(z)= {1\over 2} z\sum_{n=0}^{+\infty}{(-1)^{n}\over n!(n+1)!}\biggl
({z\over 2}\biggr )^{2n}
\label{expan}
\end{equation} 
and inverting the order of the integration and the summation, we obtain:
\begin{equation}
{f}'(s)= \sum_{n=0}^{+\infty}{(-1)^{n}\over n!(n+1)!}
\int_{0}^{+\infty}
\biggl ({z\over 2}\biggr )^{2n} e^{-s^{2}z^{2}}dz
\label{f9}
\end{equation} 
If we now introduce the variable $t=s^{2}z^{2}$, equation (\ref{f9}) becomes
\begin{equation}
{f}'(s)= \sum_{n=0}^{+\infty}{(-1)^{n}\over n!(n+1)!}{1\over (2s)^{2n+1}}
\int_{0}^{+\infty}
t^{n-{1\over 2}} e^{-t}dt
\label{f10}
\end{equation} 
or, alternatively
\begin{equation}
{f}'(s)=\sum_{n=0}^{+\infty}{(-1)^{n}\over n!(n+1)!}{1\over (2s)^{2n+1}}
\Gamma\biggl (n+{1\over 2}\biggr )
\label{f11}
\end{equation}
Integrating this series term by term, we get
\begin{equation}
{f}(s)=K+{\sqrt{\pi}\over 2}\ln s-{1\over
4}\sum_{n=1}^{+\infty}{(-1)^{n}\over n!(n+1)!}{1\over n}{1\over (2s)^{2n}}
\Gamma\biggl (n+{1\over 2}\biggr )
\label{f12}
\end{equation}
where $K$ is a constant. Numerically, we find $K=1.41548...$.
Therefore, to leading order in $s$, we can write
\begin{equation}
{f}(s)=1.41548...+{\sqrt{\pi}\over 2}\ln s \qquad (s\rightarrow +\infty)
\label{f13}
\end{equation}

\section{An alternative derivation of equation (\ref{N6})}
\label{sec_alt}

The average velocity occurring in $O$ is given by the expression
\begin{equation}
\langle {\bf V}_{0}\rangle =\int W({\bf V}_{0},{\bf V}_{1}){\bf V}_{0}d^{2}{\bf
V}_{0}d^{2}{\bf V}_{1}
\label{qQ1}
\end{equation}
Using equation (\ref{W10}) and integrating by parts, we obtain
\begin{equation}
\langle {\bf V}_{0}\rangle =-i {\partial A\over\partial {\mb\rho}}({\bf 0},{\bf
0})=i n{\partial C\over\partial {\mb\rho}}({\bf 0},{\bf 0}) 
\label{qQ2}
\end{equation}
Similarly,
\begin{equation}
\langle {\bf V}_{1}\rangle =-i {\partial A\over\partial {\mb\sigma}}({\bf
0},{\bf 0})=i n{\partial C\over\partial {\mb\sigma}}({\bf 0},{\bf 0}) 
\label{qQ3}
\end{equation}

The same method applied to the correlation function
\begin{equation}
\langle {\bf V}_{0} {\bf V}_{1}\rangle =\int W({\bf V}_{0},{\bf V}_{1})
{\bf V}_{0}{\bf V}_{1}  d^{2}{\bf V}_{0}d^{2}{\bf V}_{1}
\label{qQ4}
\end{equation}
yields
\begin{equation}
\langle {\bf V}_{0} {\bf V}_{1}\rangle =-{\partial^{2}A\over\partial
{\mb\rho}\partial {\mb\sigma}}({\bf 0},{\bf 0})=-n^{2}{\partial C\over\partial
{\mb\rho}}({\bf 0},{\bf 0}){\partial C\over\partial {\mb\sigma}}({\bf 0},{\bf
0})+n{\partial^{2}C\over\partial {\mb\rho}\partial {\mb\sigma}}({\bf 0},{\bf 0})
\label{qQ5}
\end{equation}
Hence, we can write
\begin{equation}
\langle {\bf V}_{0} {\bf V}_{1}\rangle =\langle {\bf V}_{0}\rangle \langle {\bf
V}_{1}\rangle+n{\partial^{2}C\over\partial {\mb\rho}\partial {\mb\sigma}}({\bf
0},{\bf 0})      
\label{qQ6}
\end{equation}
Since $\langle {\bf V}_{0}\rangle={\bf 0}$, there remains
\begin{equation}
\langle {\bf V}_{0} {\bf V}_{1}\rangle =n{\partial^{2}C\over\partial
{\mb\rho}\partial {\mb\sigma}}({\bf 0},{\bf 0})      
\label{qQ6bis}
\end{equation}

Substituting explicitly for $C({\mb\rho},{\mb\sigma})$ from equation
(\ref{C11}) in equations (\ref{qQ3}) and (\ref{qQ6bis}), we find
\begin{equation}
\langle {\bf V}_{1}\rangle =-{n\gamma\over 2\pi}\int { ({\bf r}-{\bf
r}_{1})_{\perp}\over |{\bf r}-{\bf r}_{1}|^{2} }d^{2}{\bf r}
\label{qQ7}
\end{equation}
and
\begin{equation}
\langle {\bf {V}}_{0} {\bf { V}}_{1}  \rangle = {n\gamma^{2}\over 4\pi^{2}}\int
{ {\bf r}_{\perp}\over r^{2}}{({\bf r}-{\bf r}_{1})_{\perp}\over  |{\bf r}-{\bf
r}_{1}|^{2}} d^{2}{\bf r}
\label{qQ8}
\end{equation}

By part integration, the first integral can be rewritten
\begin{equation}
\langle {\bf V}_{1}\rangle =-{n\gamma\over 2\pi}{\hat {\bf z}}\wedge \oint \ln
|{\bf r}-{\bf r}_{1}|{\hat{{\bf e}}_{r}}dl
\label{qQ9}
\end{equation}
where ${\hat {\bf z}}$ is a unit vector normal to the flow and ${\hat{{\bf e}}_{r}}$ is a unit vector normal to the disk boundary at $r=R$. Using the expansion (\ref{id11}) for $\ln |{\bf r}-{\bf r}_{1}|$, we obtain easily
\begin{equation}
\langle {\bf V}_{1}\rangle ={1\over 2}n\gamma {\bf r}_{1\perp}
\label{qQ10}
\end{equation}
which corresponds to the solid rotation (\ref{M11bis}).

Integrating by parts the second integral, we get
\begin{equation}
\langle {\bf {V}}_{0} {\bf { V}}_{1}  \rangle = {n\gamma^{2}\over
4\pi^{2}}\biggl\lbrace \oint \ln |{\bf r}-{\bf r}_{1}|{{\bf r}\over
r^{2}}{\hat{\bf e}}_{r}dl-\int  2\pi\delta({\bf r})\ln |{\bf r}-{\bf r}_{1}|
d^{2}{\bf r}\biggr \rbrace
\label{qQ11}
\end{equation}
Using equation (\ref{id11}), we obtain
\begin{equation}
\langle {\bf { V}}_{0} {\bf { V}}_{1}  \rangle = {n\gamma^{2}\over 2
\pi}\ln\biggl ({R\over r_{1}}\biggr )
\label{qQ12}
\end{equation}
which coincides with formula (\ref{N6}).

Of course, the relations (\ref{qQ7}) and (\ref{qQ8}) can be obtained independently of
the formalism introduced in this article. Using equations (\ref{V2}) and
(\ref{V3}), we can write
\begin{equation}
\langle {\bf V}_{0}{\bf V}_{1}\rangle ={\gamma^{2}\over 4\pi^{2}}\biggl\langle
\sum_{i,j}{{\bf r}_{\perp i}\over r_{i}^{2}}{({\bf r}_{j}-{\bf
r}_{1})_{\perp}\over |{\bf r}_{j}-{\bf r}_{1}|^{2}}\biggr\rangle
\label{dir1}
\end{equation}
where $\langle X \rangle$ denotes the average value $\int \tau({\bf
r}_{1})...\tau({\bf r}_{N}) X d^{2}{\bf r}_{1}...d^{2}{\bf r}_{N}$ associated with a decorrelated distribution of point vortices. If we split
the sum in two parts, distinguishing the contribution of vortex pairs from the
contributions of individual vortices, we obtain:
\begin{equation}
\langle {\bf V}_{0}{\bf V}_{1}\rangle ={\gamma^{2}\over 4\pi^{2}}\biggl \langle
\sum_{i=1}^{N}{{\bf r}_{\perp i}\over r_{i}^{2}}{({\bf r}_{i}-{\bf
r}_{1})_{\perp}\over |{\bf r}_{i}-{\bf r}_{1}|^{2}}\biggr\rangle
+{\gamma^{2}\over 4\pi^{2}}\biggl \langle \sum_{i=1}^{N}\sum_{j\neq i}{{\bf
r}_{\perp i}\over r_{i}^{2}}{({\bf r}_{j}-{\bf r}_{1})_{\perp}\over |{\bf
r}_{j}-{\bf r}_{1}|^{2}}\biggr\rangle
\label{dir2}
\end{equation} 
Since all vortices are identical, we find
\begin{eqnarray}
\langle {\bf V}_{0}{\bf V}_{1}\rangle ={\gamma^{2}\over 4\pi^{2}}N\int \tau({\bf
r}) {{\bf r}_{\perp}\over r^{2}}{({\bf r}-{\bf r}_{1})_{\perp}\over |{\bf
r}-{\bf r}_{1}|^{2}}d^{2}{\bf r}\nonumber\\ +{\gamma^{2}\over 4\pi^{2}}N(N-1)\int \tau({\bf
r}) {{\bf r}_{\perp}\over r^{2}}d^{2}{\bf r}\int \tau({\bf r}){({\bf r}-{\bf
r}_{1})_{\perp}\over |{\bf r}-{\bf r}_{1}|^{2}}d^{2}{\bf r}
\label{dir3}
\end{eqnarray}  
Using $\tau({\bf r})={1\over \pi R^{2}}$ and making the approximation
$N(N-1)\simeq N^{2}$ for large $N$'s, we get
\begin{equation}
\langle {\bf V}_{0}{\bf V}_{1}\rangle ={n\gamma^{2}\over 4\pi^{2}}\int {{\bf
r}_{\perp}\over r^{2}}{({\bf r}-{\bf r}_{1})_{\perp}\over |{\bf r}-{\bf
r}_{1}|^{2}}d^{2}{\bf r} +{n^{2}\gamma^{2}\over 4\pi^{2}}\int {{\bf r}_{\perp}\over
r^{2}}d^{2}{\bf r}\int {({\bf r}-{\bf r}_{1})_{\perp}\over |{\bf r}-{\bf
r}_{1}|^{2}}d^{2}{\bf r}
\label{dir4}
\end{equation} 
The second term is just the product $\langle {\bf V}_{0}\rangle \langle {\bf V}_{1}\rangle$ which is equal to zero since $\langle {\bf V}_{0}\rangle ={\bf 0}$. Therefore, equation (\ref{dir4}) returns our previous expression (\ref{qQ8}). Accordingly, the formula (\ref{qQ12}) for the velocity autocorrelation function can be derived very simply, independantly from the general formalism introduced in the article.  Note, in contrast, that  $\langle
V_{1 ||}\rangle$ and $\langle {\bf V}_{0} {\bf V}_{1}/V_{0}^{2}\rangle$ cannot be obtained by this method. Therefore, the general
theory developed in sections \ref{sec_V1perpr1} and \ref{sec_foncnorm} is necessary to compute these more complicated quantities.

\end{document}